\documentclass{aa}
\usepackage{times,graphics,rotate,psfig,epsfig}

\newcommand{\HeI} {He\,{\sc i}}
\newcommand{\HeII} {He\,{\sc ii}}

\newcommand{\NIII} {N\,{\sc iii}}
\newcommand{\NIV} {N\,{\sc iv}}
\newcommand{\NV} {N\,{\sc v}}

\newcommand{\CIII} {C\,{\sc iii}}
\newcommand{\CIV} {C\,{\sc iv}}

\newcommand{\OII} {O\,{\sc ii}}

\newcommand{\OV} {O\,{\sc v}}
\newcommand{\SiIII} {Si\,{\sc iii}}

\newcommand{\SiIV} {Si\,{\sc iv}}

\newcommand{\teff}{$T_{\rm eff}$}

\newcommand{\g} {$\log g$}

\newcommand{\eps} {$\epsilon$}
\newcommand{\Vr} {$V_{\rm r}${\thinspace}sin{\thinspace}$i$}

\newcommand{\Vi} {$V_{\infty}$}
\newcommand{\lMp} {log \.M}

\newcommand{\lMWM} {log (\.M $V_{\infty} R^{\rm 0.5}$)}

\newcommand{\Hd} {H$_{\rm \delta}$}
\newcommand{\Hg} {H$_{\rm \gamma}$}
\newcommand{\Hb} {H$_{\rm \beta}$}
\newcommand{\Ha} {H$_{\rm \alpha}$}
\newcommand{\Rsun} {$R_{\odot}$}
\newcommand{\Msun} {${\rm M}_{\odot}$}
\newcommand{\R} {$R/R_{\odot}$}

\newcommand{\lL} {$\log (L/L_{\odot})$}

\newcommand{\lR} {$\log (R/R_{\odot})$}
\newcommand{\lM} {$\log (M/M_{\odot})$}

\newcommand{\Ms} {$M_{\rm s}$}

\newcommand{\Mo} {$M_{\rm 0}$}
\newcommand{\Mev} {$M_{\rm ev}$}

\def\rarrow{\rightarrow}
\def\dd{{\rm d}}
\def\<<{{\ll}}
\def\>>{{\gg}}
\def\=={{\equiv}}

\def\inf{{\infty}}

\def\vrad{v_{\rm rad}}

\def\v0{v_{0}}

\def\ie{{i.e., }}

\def\Rstar{R_{\ast}}

\def\Rsuna{\rm{R_{\odot}}} 
\def\Msuna{\rm{M_{\odot}}} 

\def\vinf {v_\inf}
\def\Mdot{{\dot M}}

\def\taue{\tau_e}

\def\se{s_{\rm e}}
\def\Ihe{I_{\rm He}}

\def\beq{\begin{equation}}
\def\eeq{\end{equation}}
\def\beqa{\begin{eqnarray}}
\def\eeqa{\end{eqnarray}}

\def\xne{n_{\rm e}}
\begin{document}
\thesaurus{07(08.01.3; 08.05.1; 08.05.3; 08.06.3; 08.13.2)}
\title{Fundamental parameters of Galactic luminous OB stars\\
IV. The upper HR diagram\thanks
{The INT is operated on the island of 
La Palma by the RGO in the Spanish Obervatorio de El Roque de los Muchachos 
of the Instituto de Astrof\'\i sica de Canarias.}}
\author{A. ~Herrero\inst{1,2}, J. Puls\inst{3} \and M.R. ~Villamariz\inst{1}}
\offprints{A. ~Herrero}
\mail{ahd@ll.iac.es}
\institute{Instituto de Astrof\'\i sica de Canarias, E-38200 La Laguna, 
Tenerife, Spain
\and
Departamento de Astrof\'{\i}sica, Universidad de La Laguna,
Avda. Astrof\'{\i}sico Francisco S\'anchez, s/n, E-38071 La Laguna, Spain
\and
Universit\"ats-Sternwarte M\"unchen, Scheinerstr. 1, D-81679 M\"unchen, 
Germany}
\titlerunning{The Upper HR Diagram}
\date{15 June 1999; accepted date}
\maketitle
\begin{abstract}

We present observations and analyses of seven Galactic O stars of type
O6 and earlier.  The analyses are carried out using NLTE
plane--parallel, hydrostatic models as well as NLTE spherical models
with mass-loss.  With detailed calculations for the former and
simulations for the latter, it is shown that the flux blocking due to
UV metal lines is important for these objects, in agreement with
previous studies, and the way the mechanism operates is explained. We
find that the plane--parallel, hydrostatic unblanketed model
atmospheres have increasing difficulties in fitting the early-type
spectra of massive stars, and for 50\,000~K and above a fit seems to
be impossible. The gravities derived are relatively low even for the
luminosity class V stars. These objects also show the mass discrepancy
found in earlier studies, indicating that sphericity and mass-loss are
important, even at their higher gravities. We then perform an analysis
using spherical models with mass-loss. It is found that gravities
should be increased by 0.1--0.25 dex, reducing, but not solving, the
mass discrepancy.  We show that spectroscopic masses are in better
agreement with the theory of radiatively driven winds than
evolutionary masses are.  A helium abundance larger than solar is also
obtained for most objects.\\  

Some additional effects (partly related to present approximations)
that have an influence in our analyses are studied. It is found that
\HeII\,$\lambda$4200 is less sensitive to details of the model
calculations than \HeII\,$\lambda$4541 and thus it is preferred for
temperature determinations, with the consequence of lower effective
temperatures. It is shown that the fits to \HeII\,$\lambda$4686 are
improved when the upward rates of the \HeII\, resonance lines are
reduced (with respect to the conventional treatment adequate for lines
formed in expanding atmospheres), either by setting them in detailed
balance or by artificially adding extra opacity sources that simulate
line blocking. The \HeII\, blend with \Ha\, is also affected.\\  
Some stars of our sample have such high mass--loss rates that the
derivation of gravities from the wings of Balmer lines, in particular
\Hg , becomes doubtful.  For the most extreme objects, the mass--loss
rates needed to fit \Ha~ are different from those needed to fit \Hg ,
by a maximum factor of two.\\

From the point of view of individual stars, we have analysed some of
the most massive and luminous stars in the Milky Way. According to our
analysis, three of them (Cyg OB2 $\#7$, HD 15\,570 and HD\,15\,558)
have particularly large initial masses, close to or in excess of 100
\Msun . Finally, the least luminous object in our sample, HD 5\,689,
could have been erroneously assigned to Cas OB7 and might be a runaway
star.
\end{abstract}

\section{Introduction}

Massive hot stars are relevant objects in many areas of
astrophysics. They are the main sources of photoionization of their
surrounding interstellar medium, and they contribute to the chemical
and dynamical evolution of the host galaxy through their stellar winds
and supernova explosions.  They are also the precursors of Wolf--Rayet
stars and LBVs and constitute excellent probes of evolutionary models.
For these reasons, advances in our understanding of these objects have
an impact on other fields of astrophysics.

In spite of all this, few attempts have been made to analyse massive O
stars quantitatively. The most comprehensive study up to now has been
that of Herrero et al. (\cite{h92}, hereafter Paper I), who analysed
24 stars from spectral types B0 to O5.  Detailed quantitative analyses
of massive stars of early spectral type have been even more
scarce. Conti \& Frost (\cite{conf77}) first made a systematic
analysis of the earliest spectral types, comparing with theoretical
predictions by Auer \& Mihalas (\cite{aum72}), but only for gravities
of \g = 4.0. Later, Kudritzki (\cite{kud80}) studied HD~93250, an O3~V
star, using models similar to those of Auer \& Mihalas (\cite{aum72}),
and Kudritzki et al. (\cite{kud83}) studied the spectrum of $\zeta$
Pup and showed that the star has a low gravity (of the order of \g =
3.5) and a high helium abundance (\eps = $N$(He)/($N$(H)+$N$(He))=
0.14). Kudritzki \& Hummer (\cite{kh90}) list 15 stars earlier than
spectral type O6 (only three classified as supergiants), with
parameters determined using the same methods.  Puls et
al. (\cite{puls96}) list 22 stars earlier than O6 in the Milky Way and
the Magellanic Clouds, still taking advantage of the optical analysis,
but already incorporating the effects of sphericity and mass-loss.
Pauldrach et al. (\cite{pau93}) analysed Melnick 42 and $\zeta$ Pup,
using only the UV spectrum, and Taresch et al.  (\cite{tar97}) made a
very detailed analysis of HD 93\,129A using the UV and FUV
spectrum. Finally, de Koter (\cite{dK98}) has used the UV \OV~ line at
1371 \AA~ to determine the temperature of very hot stars in R136a.

All these studies have revealed a number of problems in the analyses
of these stars, related to our incomplete understanding of these
objects.  In Paper I, the so--called helium and mass discrepancies,
already present in the previous literature (see Kudritzki et
al. \cite{kud83}; Voels et al. \cite{vo89}; Groenewegen et
al. \cite{groe89}; Herrero et al. \cite{h90}) were shown to be
systematic. These refer to the discrepancy in the values of the
stellar mass and the photospheric helium abundance obtained from the
analysis of the spectrum using state-of-the-art model atmospheres and
evolutionary models. The explanation of these discrepancies is still
unclear (for recent working directions, see Howarth \cite{how98}).

In Paper I we already noted the correlation between the mass
discrepancy and the distance of the star to the Eddington limit,
indicating that the plane--parallel geometry and the hydrostatic
equilibrium assumption could be the reason for the low stellar masses
derived. However, the use of wind techniques in the same work already
indicated that the discrepancy could be reduced, but not solved, by
including mass-loss and sphericity effects. This was later confirmed
in an analysis of HDE~226\,868 (Herrero et al. \cite{h95}), where the
authors used Unified Model atmospheres to determine the mass of this
star, the optical counterpart of Cygnus X--1, by combining the
spectroscopic analysis with the orbital data.  A similar result has
been obtained by Israelian et al. (\cite{isr99}) in an analysis of
HD~188\,209.  The inclusion of mass-loss and sphericity did not seem
to have any effect on the helium discrepancy either (Herrero et al.,
\cite{h95}) in spite of the variations that strong winds could
introduce in the helium profiles as compared to static,
plane--parallel atmospheres (Schaerer \& Schmutz \cite{ss94}). One of
the reasons for this weak influence was the fact that the study in
Paper I was limited to spectral types of O5 or later, because it was
found that above 40\,000~K the neutral helium singlet and triplet
lines gave different stellar parameters. Herrero (\cite{h94})
suggested that this was due to the neglect of the so-called {\em
line-blocking}, the UV background opacity due to metal lines, during
the line formation calculations.  Also the inclusion of
microturbulence in these calculations can reduce this difference
(McErlean \cite{mce98}; Smith \& Howarth \cite{sh98}; 
Villamariz \& Herrero, in prep.).

In this paper we study a few stars of early spectral type in an
attempt to cover several objectives. First, we would like to extend
our sample from Paper I towards earlier spectral types and thus cover
the whole region of interest in the HR diagram with plane--parallel
analyses to see their complete behaviour. Although the plane--parallel
models will have difficulties in explaining even the optical spectrum
of these very hot stars (see Sect.~\ref{anap}), this first step is
needed for the subsequent application of more sophisticated models,
which will use the experience gained and the parameters obtained as
input.  An analysis of line-blocking effects is mandatory here, as it
has been seen to have an influence for the higher temperatures.  

Then, we will repeat the analyses using spherical models with
mass-loss, and present the study of some effects that both can
influence the determination of stellar parameters and that
will help us to gain new insight into the physics of these
stars. Having these parameters determined, we can try to establish
conclusions with respect to the use of different model atmospheres and
techniques for the analysis of very early spectral types. We will
compare the results of the spherical models with mass-loss with those
from plane--parallel, hydrostatic model atmospheres 
as well as with the results obtained using the somewhat approximate
technique employed by Puls et al. (\cite{puls96}).

In Sect.~\ref{secobs} we present the observations. The effects of
line-blocking in plane--parallel models are treated in
Sect.~\ref{lblock} and the spectral description and plane--parallel
analysis are considered in Sect.~\ref{anap}. Sect.~\ref{anas} shows
the analyses performed with spherical models with mass-loss, while
Sect.~\ref{numef} contains a qualitative study of some effects of
interest which explain some difficulties found in the preceding
section.  Then we present our discussion (Sect.~\ref{discu}) and
conclusions (Sect.~\ref{conc}).

\section{The observations}
\label{secobs}

The observations were carried out with the 2.5~m Isaac Newton
Telescope at the Roque de los Muchachos Observatory on La Palma during
two different observational runs, in 1991 September and 1992 August.
We have observed the spectral region between 4000 and 5000 \AA , and
the region around \Ha .  The Intermediate Dispersion Spectrograph was
used with the H 2400 B grid in the blue and the H 1800 V grid in the
red, attached to the 235 mm camera, which resulted in spectral
resolutions of 0.6 and 0.8 \AA, respectively, measured on the Cu--Ar
arc. The main difference between both runs was the size of the CCD
detector, with a larger wavelength coverage per exposure (400 \AA~
instead of 220 \AA~ in the blue) during the second run.
Table~\ref{obs} gives the stellar identification (usually the HD
number), star name, OB association to which the star belongs, spectral
type and the night in which the stellar spectrum was obtained.

\begin{table*}
\label{obs}
\caption[ ]{Journal of observations. September dates refer to 1991, and
August dates to 1992. The night of 9/10 August was dedicated to \Ha~
observations. All spectral classifications 
are taken from Walborn (\protect\cite{wal72}, \protect\cite{wal73}), 
except that of
HD\,5\,689, which is taken from Garmany \& Vacca (\protect\cite{gv91}). 
Properties
indicated in the last column have been taken from Mason et al. 
(\protect\cite{mas98})}
\begin{flushleft}
\begin{tabular}{lllllll} 
\hline
Order & Identification    & Spectral Type & Name & Assoc. & Observing nights
& Notes \\
\hline
1 & Cyg OB2 \#7 & O3 If$^{*}$  &     & Cyg OB2 & 6/7, 9/10 Aug  & \\
2 & HD 15\,570 & O4 If$^{+}$   &     & Cas OB6 & 6/7/8, 9/10 Aug & \\
3 & HD 15\,629 & O5 V((f)) &     & Cas OB6 & 28/29 Sep; 7/8, 9/10 Aug &     \\
4 & HD 15\,558 & O5 III(f)    &     & Cas OB6 & 28/29 Sep; 9/10 Aug & binary \\
5 & HD 14\,947 & O5 If$^{+}$     &     & Per OB1 & 28/29 Sep; 9/10 Aug &     \\
6 & HD 210\,839& O6 I(n)fp &$\lambda$ Cep & Cep OB2 & 28/29 Sep ; 9/10 Aug & runaway \\
7 & HD 5\,689  & O6 V       &     & Cas OB7 & 8/9, 9/10 Aug &  \\
\end{tabular}
\end{flushleft}
\end{table*}

The reduction of the data was done following standard procedures. We
used both IRAF\footnote{The IRAF package is distributed by the
National Optical Astronomy Observatories, which is operated by the
Association of Universities for Research in Astronomy, Inc., under
contract with the National Science Foundation}and own software
developed in IDL. The latter was also used for the spectral analysis.
The S/N ratio of the reduced spectra depends on the spectral range,
but is usually about 200 in the \Hg~region.

\section{Line blocking in plane--parallel models}
\label{lblock}

In Paper I we found that the \HeI~ singlet and triplet lines of O
stars hotter than 40\,000~K indicated different stellar parameters
when used for the determination of stellar temperatures.  Following
the results of Pauldrach et al. (\cite{pau93}), who stressed the
importance of line blocking for the wind ionization structure and
emergent flux, Herrero (\cite{h94}) showed that this difference
is considerably reduced when UV metal line opacity is included.  The
author attributed this to the different effect of the modified UV
radiation field on the He~{\sc i} occupation numbers. However, no
further proof was given there.

Here we want to discuss in a more detailed way the effect of this
line blocking, as it is important for the analysis of the stars
listed in Table~\ref{obs}.

We have included the line list of Pauldrach et al. (\cite{pau93})
between 228 and 912 \AA . In this region the ionization of \HeI~ and H
takes place.  Thus, we can expect an effect on the occupation numbers
of these two ionization stages. The line list comprises 
roughly 14\,000 lines from
26 different elements and 137 ionization stages taken into account in
the stellar wind NLTE calculations by Pauldrach et
al. (\cite{pau93}). We calculate their occupation numbers in
LTE. Although this is a rough approximation, we expect that its
inclusion will already give us the major part of the effect.

We will use a model at \teff\ = 40\,000 K, \g\ = 3.40 and \eps\ = 0.09
to illustrate the effects of line blocking. In Fig. 1 we plot the
emergent flux between 228 and 912 \AA . As we can see, a considerable
fraction of the flux has been blocked. This energy will appear at
other wavelengths. However, as most of the flux escapes between 912
and 2\,000 \AA , this has no real influence redward of 912 \AA . It
has to be pointed out that Fig. 1 is only illustrative. It is not the
emergent flux which is important here, but the mean intensity of the
radiation field at the depths where the \HeI~ continuum becomes
optically thin. The effect, however, is similar.

\begin{figure*}
{\psfig{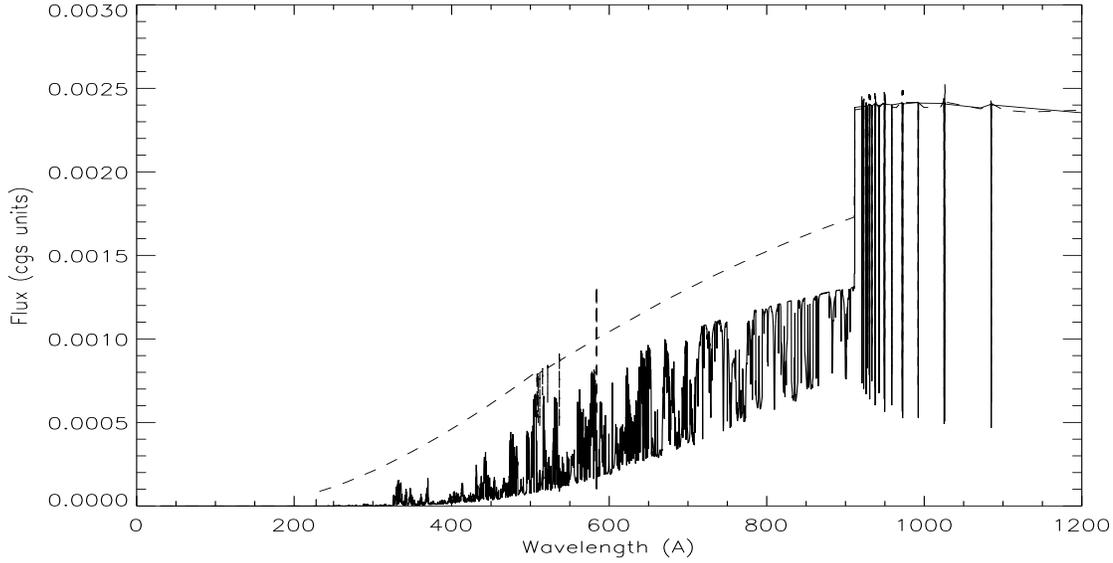}}
\caption[]{The Eddington flux in the region between 228 and 1200 \AA , with and
without approximate line blocking, for the model at \teff\ = 40\,000 K, \g\ =
3.40 and \eps\ = 0.09.}
\end{figure*}

\begin{figure*}
{\psfig{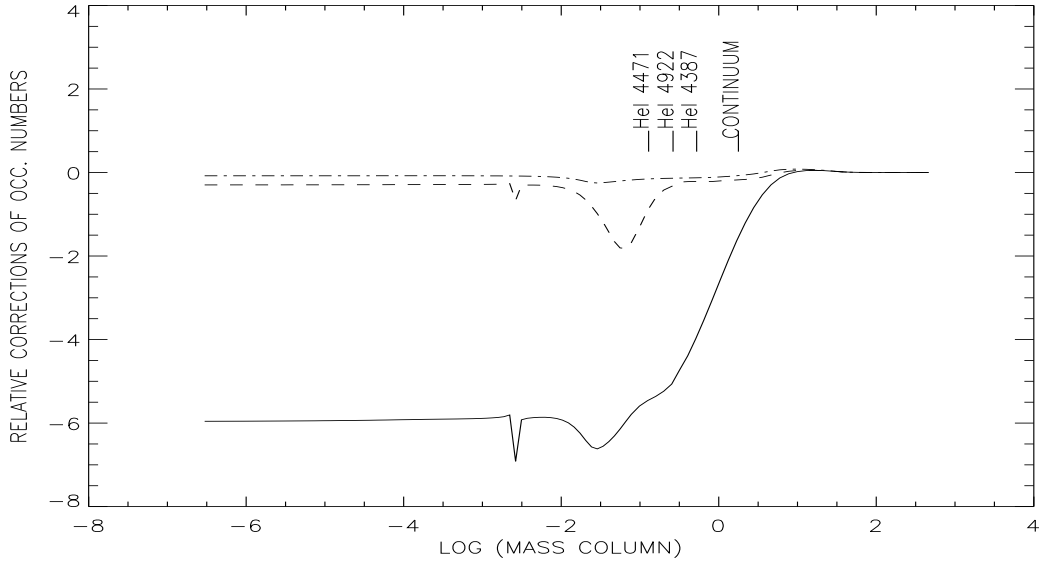}}
\caption[]{The changes in the occupation numbers of the ground level
of \HeI~ (full line), the lower level of the singlet lines 
\HeI~ $\lambda\lambda$4387, 4922 (2p $^1P^0$, dashed
line) and the ground level of the triplet line \HeI~ $\lambda$4471
(2p$^3P^0$, dashed--dotted),
and the formation depths of the line centers and the continuum.
Negative numbers mean that the atomic level populations calculated with
line-blocking are larger.
The peak near log $m$ = --2.5 is a numerical artefact in the
convergence of the model without line blocking and does not 
appear in other models. The model parameters are the same as in Fig. 1.}
\end{figure*}

The most important effect is produced in the occupation numbers of the
ground level of \HeI . As a consequence of the flux blocking, the
ionization from this level (its ionization edge lies at 504 \AA ) is
considerably reduced, and its population largely increases (see
Fig. 2). This does not have any noticeable effect on other ionization
stages, since \HeI~ is very scarcely populated. The lower level of the
\HeI~ $\lambda\lambda$4387, 4922 singlet lines, the 2p $^1P^0$ level,
is directly connected to the ground level (which also belongs to the
singlet system) through a radiative transition at 584 \AA , and thus
partially follows their changes and also increases its population. On
the contrary, the lower level of the \HeI~$\lambda$4471 triplet line
(the 2p $^3P^0$ level) is only weakly or indirectly coupled to the
\HeI~ ground level.  In addition, the ground level of the triplet
system (to which 2p $^3P^0$ is strongly coupled) is dominated by its
ionization and recombination at around 2600 \AA , and it is not
affected by the line-blocking.  As a result, the behaviour of 2p
$^3P^0$ follows that of the \HeI~ ground level to a much lesser extent
than does the 2p $^1P^0$ level (as do all triplet levels compared to
their singlet counterparts). We can see this behaviour in Fig. 2,
where we have additionally marked the formation depths of the center
of the \HeI~$\lambda\lambda$ 4387, 4922 and 4471 lines and the
continuum.  We can see that the populations of these levels increase
over the whole region of formation, the changes being larger for the
2p $^1P^0$ level. In addition, the formation region of the singlet
lines is more extended than in the case without line-blocking (an
effect that is not seen in the figure). The combination of larger
changes over a larger line formation region produces the stronger
variations of the singlet lines compared to the triplet one.  

\begin{figure}[t]
{\psfig{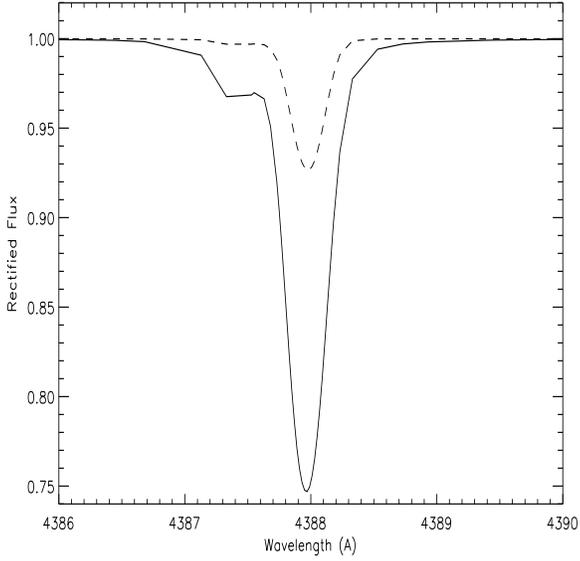}}
\caption[]{The helium line \HeI~ $\lambda$4387 with and without 
line-blocking (solid and dashed lines, respectively) for the same
model parameters as in Fig. 1. \HeI~ $\lambda$4922 behaves the same way.}
\end{figure}

\begin{figure}[t]
{\psfig{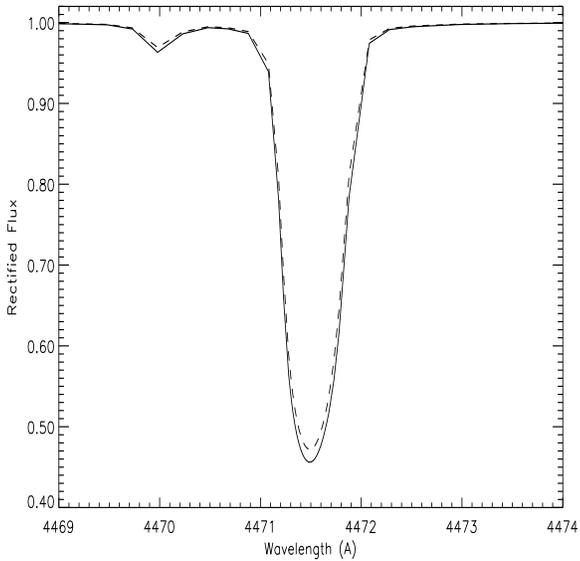}}
\caption[]{As Fig. 3, but for the \HeI~ $\lambda$4471 line.}
\end{figure}

Figs. 3 and 4 show the variations in the line profiles of He~{\sc i}~
$\lambda\lambda$4387, 4471. We see that the changes are much less in
the second one, and thus this line should be preferred for analyses at
high temperatures. At lower temperatures and gravities, however, and
due to the so-called {\em dilution effect} in this line, the use of
the singlet lines is preferable, even if the calculations are made
without line-blocking. This effect refers to the fact that the fit of
the He~{\sc i}~ $\lambda$4471 line becomes worse when going from
dwarfs to supergiants, while the rest of the lines retain a good fit
quality (see Voels et al. \cite{vo89}).  
Although still not completely clarified, Smith \& Howarth
(\cite{sh98}) recently claimed that microturbulence could be the
cause.

At lower temperatures or larger gravities, collisions play a
stronger role, and the effects of line blocking are lower. This
explains the behaviour of the corrections to the stellar parameters
found by Herrero (\cite{h94}, especially Fig.~3) when including
line-blocking, and basically
consist of obtaining lower temperatures when line-blocking is not
taken into account. The amount of the correction will depend on the
model temperature at a given gravity, being larger for higher
temperatures. 
It has to be stressed, however, that the corrections are
large only if we use the singlet lines for the temperature
determination without line--blocking. 
Using the triplet line results in a smaller temperature 
than using the singlet ones (by 500--1\,000
K at temperatures around 45\,000 K).
Gravity and helium abundance do not significantly vary, although
sometimes variations of the parameters within our typical error boxes
($\pm$0.1 in \g~ and $\pm$0.03 in helium abundance) have been adopted
in the course of the parameter determinations.

\section{Spectral analysis using plane--parallel, hydrostatic models}
\label{anap}

Fig. 5 shows the spectra of all observed stars between 4000 and 5000 \AA ,
whereas Fig. 6 shows the spectra around \Ha .  

Before performing the comparison with the theoretical model
atmospheres, we have corrected the spectra from radial velocity
displacements.  Correction for the radial velocities is crucial in the
very early type stars, because the H and He lines commonly used for
the spectral analysis may be contaminated by wind effects that fill
their red wings.  However, these corrections are particularly
difficult: the cores of strong lines may be affected by the wind and
only a few weak metal lines are present. Rotation adds a new handicap,
as it broadens lines, making them shallower and favouring blends.  In
addition the limited spectral range of each single frame limits the
number of suitable lines on individual spectra.  Thus the first
difficulty is to find a set of lines appropriate for the measurement
of the radial velocity.

\begin{figure*}
{\psfig{figure=h1606.f5,width=17.0cm,height=9.5cm}}
\caption[]{The blue stellar spectra. The relative fluxes have been
arbitrarily displaced in steps of 0.5 for the sake of clarity.
Wavelengths are in \AA.}
\end{figure*}
\begin{figure*}
{\psfig{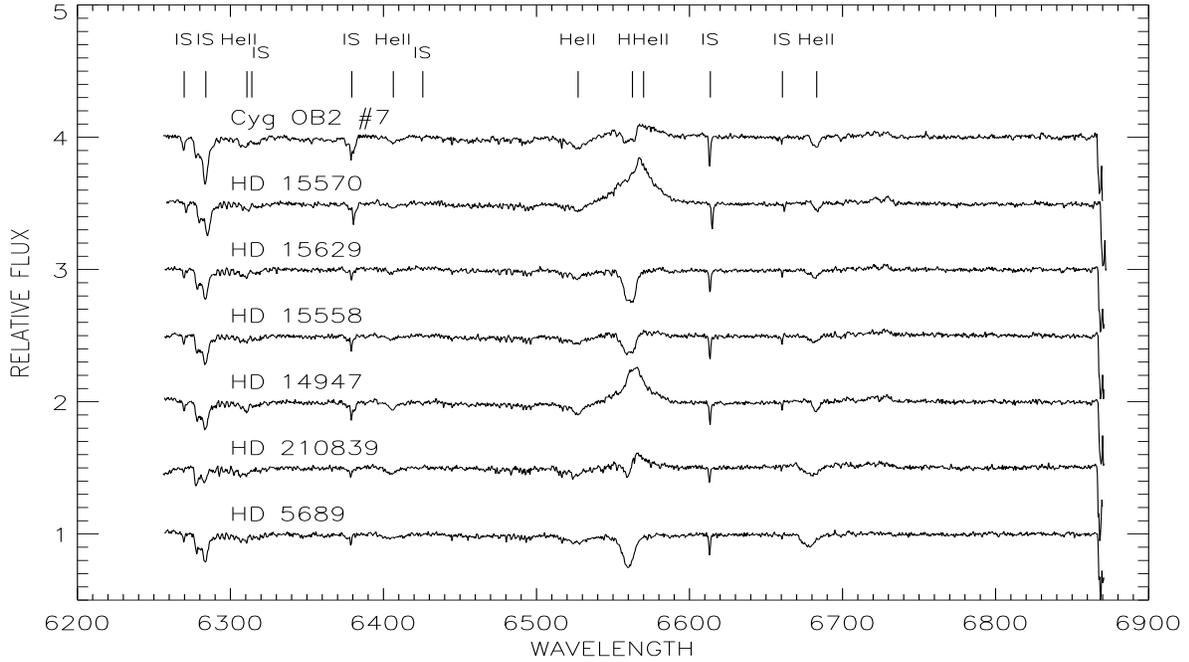}}
\caption[]{As Fig. 5, however for the red spectra}
\end{figure*}

We have discarded all H and He lines. This is already necessary, since
sometimes we can see a trend in the measured radial velocity with the
excitation potential of the line, indicating that the line cores are
formed in higher layers moving faster as the excitation potential
decreases.

Thus the radial velocity correction is based on lines of Si (\SiIV~
and \SiIII), N (\NV, \NIV~ and \NIII), O (\OII) and C
(\CIII). Sometimes, a few lines are in emission and not in
absorption. These lines are used only if they give results concordant
with the rest of the metal lines (usually, this is the case). The
typical uncertainties we obtain are $\pm$20 km s$^{-1}$, which is
about half of the spectral resolution.  This large error is due to
the existence of only a few lines, in addition with broad cores, and
reflects the dispersion in the individual values.  Within these
limits, we have sometimes displaced the fitted lines (also in the
analysis with spherical models), when it was clear that this was the
only way to obtain a good fit. Table 2 lists the
radial velocities in the kinematical Local Standard of Rest.

We then determined rotational velocities following the same procedure
as indicated in Paper I. Our values compare well with those by Penny
(\cite{penny96}) and Howarth et al. (\cite{how97}) for objects in
common (see discussion about individual objects).  Rotational
velocities are given in Table 2 together with the parameters determined
for each star. In this table, temperatures are given in thousands of
Kelvin; gravities are corrected for centrifugal force effects (see
individual discussions for model parameters); \eps~ is the helium
abundance by number with respect to the total number of H and He
atoms; $V$ is the integral of the stellar flux over $\lambda$,
weighted by the $V$-filter function of Matthews \& Sandage
(\cite{matsan63}), used to calculate stellar radii from the model
atmospheres (see Kudritzki \cite{kud80}, or Paper I); $M_V$ is the
absolute magnitude. This has been obtained using the photometry and
colours from {\it Hipparcos} for HD 15\,570, HD 15\,629, HD 15\,558,
HD 14\,947 and HD 210\,839 combined with the extinction laws and
distances from Garmany \& Stencel (\cite{gs92}). For Cyg OB2 $\#$7,
$M_V$ has been taken from Massey \& Thompson (\cite{masth91}), and for
HD 5\,689 has been taken from Garmany \& Stencel (\cite{gs92}). The
evolutionary masses have been obtained from the evolutionary tracks by
Schaller et al. (\cite{sch92}).

Except for the line-blocking, the models are the same ones used in
Paper I.  These are H/He, , plane--parallel models, in hydrostatic
and radiative equilibrium. The line fit is also made in the same
way. We first fit \Hg , obtaining the best gravity at a given \teff.
The same procedure is followed for the helium lines, at a normal He
abundance (i.e., 0.09 by number).  The locus where all lines intersect
in the log \teff -- \g~ diagram is taken as giving the stellar
parameters. If the lines do not intersect, the helium abundance is
changed. The helium abundance giving the smallest intersection region
is adopted as the one appropriate for the star.  In previous studies,
the helium lines used were \HeII~ $\lambda\lambda$ 4541, 4200 and
\HeI~ $\lambda\lambda$ 4387, 4922, whereas \HeI~ $\lambda$ 4471 was
used only for dwarfs or high-temperature stars.  As this is the case
in the present study, this is the line we use here.  In addition, for
the reasons explained before, we give it a larger weight than for the
singlet lines.

The errors are similar to those quoted in Paper I, with $\pm$1500 K in
\teff , $\pm$0.1  in \g~ and $\pm$0.03 in \eps . This produces
errors of $\pm$0.06, $\pm$0.19 and $\pm$0.22 in \lR , \lM~ and \lL ,
respectively, when adopting an uncertainty of $\pm$0.3 mag for $M_V$.

\begin{table*}
\label{param}
\caption[ ]{Parameters determined for the programme stars
using plane--parallel models. Note that
the parameters of Cyg OB2 $\#$7 and HD 15\,570 are only indicative,
and could not be determined with the plane parallel models. 
Temperatures are in thousands of Kelvin. \Ms ,
\Mev~ and \Mo~ are respectively, the present spectroscopic and 
evolutionary masses, and the initial evolutionary mass,
in solar units. The last column indicates whether we have formally a
mass discrepancy considering an error of 0.22 in log(\Ms ). \Vr~ and 
$\vrad~$ are given in km s$^{-1}$. Semicolons in $\vrad~$ indicate that
the uncertainty in this value is larger than $\pm$20 kms$^{-1}$}
\begin{flushleft}
\begin{tabular}{lccccccccccccc} 
\hline
Star & $\vrad$ & \teff~ & \g~ & \eps~ & \Vr~ & $V$ & $M_V$ & \R~ & \lL~ & \Ms~ & \Mev~ &
\Mo & M discr.?\\
\hline
Cyg OB2 \#7& +35 & 51.0 & 3.66 & 0.12 & 105 & --29.655 & --6.20 & 16.7 & 6.23 & 46.2
& 111.8 & 114 & Yes\\
HD 15\,570 & -20 & 50.0 & 3.51 & 0.15 & 105 & --29.553 & --6.73 & 22.0 & 6.44 & 51.1
& 139.1 & 142 & Yes\\
HD 15\,629 & -50 & 48.0 & 3.81 & 0.09 &  90 & --29.571 & --5.52 & 12.7 & 5.89 & 37.5
&  69.9 & 71 & Yes\\
HD 15\,558 & -35 & 46.5 & 3.71 & 0.07 & 120 & --29.512 & --6.28 & 18.5 & 6.16
& 63.8 &  91.7 & 97 & No \\
HD 14\,947 & -35: & 45.0 & 3.53 & 0.15 & 140 & --29.415 & --5.69 & 14.8 & 5.91 & 26.5
&  65.8 & 68 & Yes\\
HD 210\,839& -70 & 41.5 & 3.47 & 0.25 & 250 & --29.357 & --6.17 & 18.9 & 5.98 & 38.7
&  67.0 & 71 & Yes\\
HD 5\,689  & -65 & 40.0 & 3.57 & 0.25 & 250 & --29.337 & --4.19 &  7.7 & 5.13 &  7.8
&  30.1 & 31 & Yes\\
\end{tabular}
\end{flushleft}
\end{table*}

We now describe the line fits of each star independently, and the
spectral features, if appropriate. In Fig. 7 we show the theoretical
HR diagram for these stars, with the parameters listed in Table 2.

\begin{figure}
{\psfig{figure=h1606.f7,width=8.0cm,height=8.0cm}}
\caption[]{The programme stars on the Hertzsprung--Russell diagram
after the plane--parallel analysis. The meanings
of the symbols are those indicated in the lower left corner, together with
typical error bars. The theoretical tracks are from Schaller et al. 
(\cite{sch92}). The numbers to the left of the zero-age main sequence 
indicate the initial stellar masses in units of solar masses. $L$ is in 
solar units.}
\end{figure}

\subsection{HD 5\,689, O6 V} 
This is the only star for which we can clearly see lines of \HeI\,
other than \HeI\,$\lambda$4471.  Its line fit, shown in Fig. 8, is
obtained at \teff = 40\,000 K, \g = 3.40, \eps = 0.25 and \Vr = 250 km
s$^{-1}$. The rotational velocity, however, is uncertain, as in this
cases we only have the He lines (especially \HeI ). The given value
can be seen as an upper limit to the projected rotational velocity.
We must correct for centrifugal force by adding the term (\Vr)$^2/R$,
since the measured value is the effective gravity reduced by the
centrifugal acceleration, which has to be added here in order to
derive the masses. Corrected for centrifugal forces
the gravity in Table  2 is \g\ = 3.57. As usual for fast rotators, the
helium abundance obtained is very large. The radius, however, is very
small, and the derived spectroscopic mass is much lower than the
predicted evolutionary one. These facts reflect certain problems in
the set of parameters for this star. The star has been classified as
O6 V by Garmany \& Vacca (\cite{gv91}). This classification is not in
complete agreement with the star belonging to Cas OB7 (see Humphreys
\cite{hum78}, or Garmany \& Stencel \cite{gs92}), since the absolute
magnitude derived from the association distance corresponds to a less
luminous object than an O6 V star by one magnitude, which causes the
small radius derived (see Table 5 of Vacca et al. \cite{vacc96}). The
situation is of course much worse if, based on the low gravity, we
associate this star with a giant and adopt parameters characteristics
of a luminosity class III object.  Thus we have two possibilities: 1)
the star is a main sequence star that belongs to Cas OB7, with a
smaller radius than usual for its spectral classification, in spite of
the large rotational velocity; in this case, the low gravity is
difficult to explain, or the models are giving us completely wrong
parameters for this star; or 2) the star has parameters typical for a
luminosity class III object.  In this case, the star would probably
not belong to Cas OB7, although we have found no indications in the
literature about this possibility. Other combinations of the above
arguments are also possible, but the question that something is
non-standard with the star remains.  This case is very similar to the
one we had in Paper I for HD~24\,912, which is a known runaway
star. Using the Oort constants given by Lang (\cite{lang80}), we
obtain a peculiar velocity of 39 km s$^{-1}$ for HD\,5689, which,
given our uncertainties, does
not allow us to decide clearly whether it is a runaway or not, if we
adopt the conventional limit of 30 km s$^{-1}$ for runaway stars.

\begin{figure}
{\epsfig{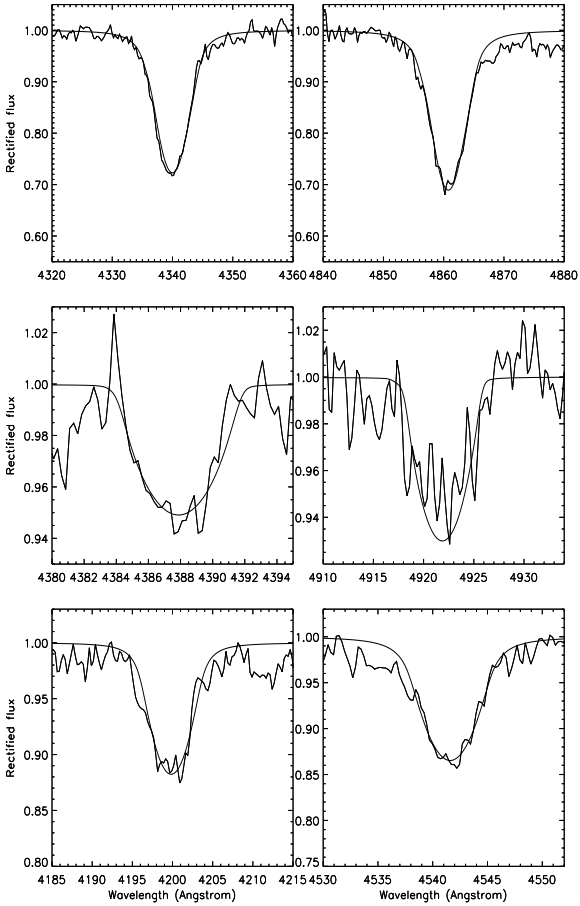}}
\caption[]{The fit to the spectral lines in HD 5\,689 using the 
plane--parallel, hydrostatic models, with the parameters given in Table 2.
From top to bottom and left to right: \Hg , \Hb , \HeI~ 
$\lambda\lambda$4387, 4922, \HeII~ $\lambda\lambda$4200, 4541.}
\end{figure}

\subsection{HD 210\,839, O6 I(n)fp} 
This is $\lambda$ Cep, a well known fast rotator.  The line fit (see
Fig. 9) is obtained at \teff\ = 41\,500, \g\ = 3.40, \eps\ = 0.25 and
\Vr = 250 km s$^{-1}$.  The rotational velocity is again uncertain and
could be lower, as we suggest later.  This is also indicated by the
values found by Penny (\cite{penny96}; 214 km s$^{-1}$) and Howarth et
al. (\cite{how97}; 219 km s$^{-1}$).  With the centrifugal force
correction, \g~ increases to 3.47 (see Table 2).  The line fit is the
best agreement we could find between all the He lines.  The parameters
are very similar to those of HD 5\,689, (except that now the mass
discrepancy is only the usual factor 2) but the spectrum shows
important differences.  $\lambda$ Cep displays Of features, and has
strong emission in \Ha . This has to be attributed to the large
difference in luminosity (see Fig. 7 and Table 2).  In Fig. 7, HD
210\,839 occupies the position of an evolved luminous star, already
within the instability strip predicted by Kiriakidis et
al. (\cite{kk93}). 
In agreement with this location, $\lambda$ Cep is known to show
profile variations due to non-radial pulsations, which however are
below the accuracy of our optical spectra (cf. Fullerton et
al. \cite{fgb96}, de Jong et al. \cite{dJo99}).  (Note, that we have seen
temporal variations in the \Ha~ profile which was observed several
times. In contrast, however, two spectra taken in the region from 4200
to 4600 \AA~ showed no significant variations in the lines used for
the spectral analysis.)
Finally, let us remark that the IR spectrum of $\lambda$ Cep has 
been analysed recently by Najarro et al. (\cite{naj98}), who obtained
parameters very similar to ours (except in the temperature, for which they
find a value lower by 4\,000 K, see discussion).

\begin{figure}
{\epsfig{figure=h1606.f9,width=8.0cm,height=12.0cm}}
\caption[]{As Fig. 8, for HD 210\,839. Here we show \HeI~
$\lambda$ 4471 instead of \HeI~ $\lambda$ 4922.}
\end{figure}

\subsection{HD 14\,947, O5 If$^+$} 
The spectral lines of this star are fitted at \teff\ = 45\,000 K, \g\
= 3.50, \eps\ = 0.15 and \Vr\ = 140 km s$^{-1}$, which agrees with the
value of 133 km s$^{-1}$ given by Penny (\cite{penny96}) and Howarth
et al. (\cite{how97}). The spectral line fits are shown in
Fig. 10. The star is an extreme Of (without being a transition object,
see Conti et al. \cite{con95}), with a large emission in
\NIII~$\lambda$4630--40 and \HeII~ $\lambda$4686, and also in \Ha .
We can also begin to see \SiIV\,$\lambda$4116 in emission and the
\NV\,$\lambda\lambda$4604, 4620 lines in absorption. The line fit,
shown in Fig. 10, can be considered as acceptable. The fit of the
\HeI~ $\lambda$4922 line indicates that the predicted line is a little
too strong as compared to the observations, but the difference is very
small compared to variations of the line within the error box.  The He
abundance is not as large as for the two previous objects, but is
still larger than solar. The spectroscopic mass is more than a factor
of two smaller than the evolutionary one.

\begin{figure}
{\epsfig{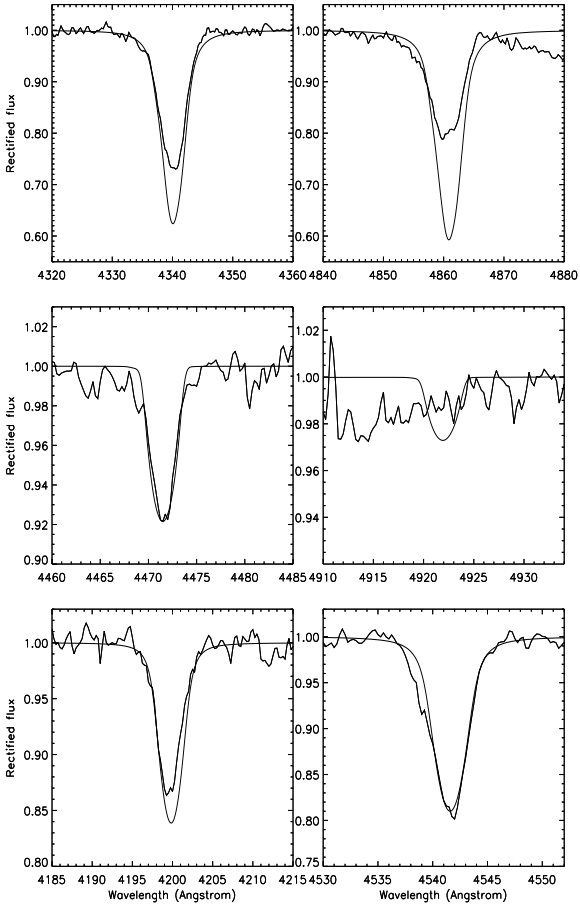}}
\caption[]{As Fig. 8, for HD 14\,947. Here we show \HeI~ $\lambda$ 
4471 instead of \HeI~ $\lambda$ 4387.}
\end{figure}

\subsection{HD 15\,558, O5 III(f)} 
This star is a binary, but we expect the spectrum not to be
contaminated, as it is a single component in a well separated system
(Mason et al. \cite{mas98}).  The best line fit, shown in Fig. 11, is
obtained at \teff\ = 46\,500 K, \g\ = 3.70, \eps\ = 0.07 and \Vr\ =
120 km s$^{-1}$.  The rotational velocity value agrees with the 123 km
s$^{-1}$ of Howarth et al.  (\cite{how97}) although it departs
slightly from the one given by Penny (\cite{penny96}) of 147 km
s$^{-1}$.  The Of features are weaker than in the previous star, as is
the emission in \SiIV\,$\lambda$4116, but note that the gravity is now
larger.  The line fit with hydrostatic models is very difficult. Only
the wings of \Hg~ and \Hb~ are fitted, together with \HeII~
$\lambda$4541. The singlet \HeI~ lines are again too weak and noisy,
but in this case also \HeI~ $\lambda$4471 cannot be fitted. We cannot
simply attribute the distortion in the blue wing of \HeI~
$\lambda$4471 to binarity. If this were the case, we should see the
singlet lines too (if the companion is relatively cool) or distortions
in the \HeII~ lines (if the companion is relatively hot).  The lack of
a good fit in any \HeI~ line makes the parameter determination much
less accurate.  From the plane--parallel hydrostatic models this is
the most massive star, and in fact the mass discrepancy is
comparatively low, the difference between the spectroscopic and
evolutionary masses being only 30$\%$.

\begin{figure}
{\epsfig{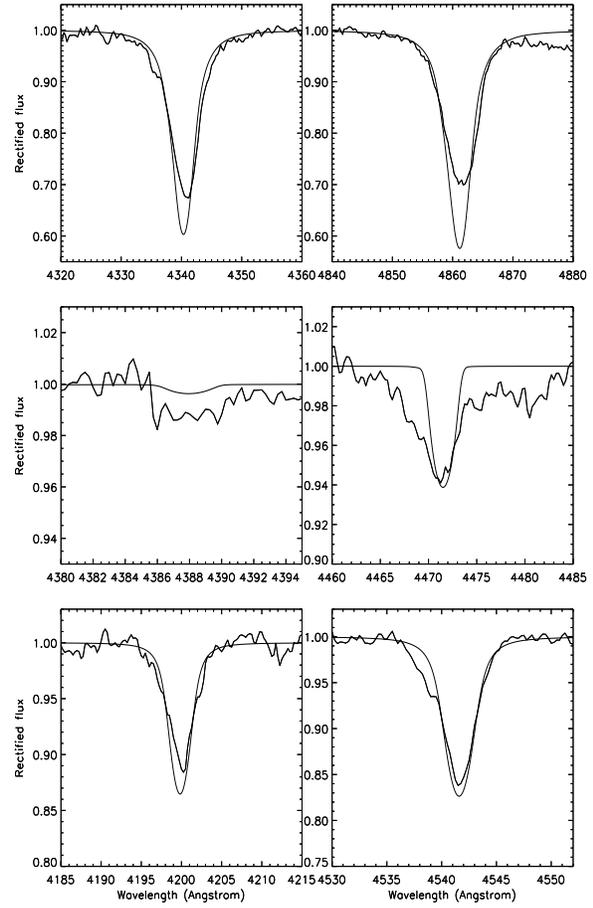}}
\caption[]{As Fig. 9, for HD 15\,558.}
\end{figure}

\subsection{HD 15\,629, O5 V((f))} 
The best line fit for this star is obtained at \teff\ = 48\,000 K, \g\
= 3.80, \eps\ = 0.09, with \Vr\ = 90 km s$^{-1}$.  This last value
agrees within $\pm$1 km s$^{-1}$ with the values quoted by Penny
(\cite{penny96}) and Howarth et al. (\cite{how97}).  In spite of the
high temperature, the larger gravity allows us the calculation of
hydrostatic models. The line fit is shown in Fig. 12. We can see that
the fit is good for the wings of \Hg~ and \Hb , and for \HeI~
$\lambda$4471 and \HeII~ $\lambda$4541, but it is bad for \HeI~
$\lambda$4387, and also for \HeI~$\lambda$4922, that is
not shown in Fig. 12. We have given much less weight to these
two lines, however, because they are very weak and noisy.  In the case
of \HeI~ $\lambda$4922, the local continuum has also been placed too
low. Correction of this would bring the calculated line into agreement
with the observations. More worrying is the lack of fit in the \HeII~
$\lambda$4200 line. The fit of this line is bad, and comparable to
that of the \HeII~ $\lambda$4686 line, for which we expect a bad fit
with hydrostatic models. For this reason we preferred to give more
weight to the fit of \HeII~ $\lambda$4541 than to \HeII~
$\lambda$4200. In spite of the large gravity and the luminosity class
V, this star also shows the mass discrepancy, which was not the case
for less luminous stars in Paper I, where we found no significant mass
discrepancy for stars of high gravities.  This indicates the increasing role
of radiation pressure.

It is interesting to note
the case of HD\,15\,629, HD\,14\,947 and
HD\,210\,839. Within our error bars, these stars could represent 
different evolutionary stages of a star of initially around 70 \Msun\,
following standard evolutionary tracks with mass loss but without rotation.
Within this scenario, it is impossible to explain the higher
He abundances of the two cooler stars. This can be an indication that 
rotation plays a strong role in
stellar evolution, since there is no other known mechanism that might
substantially modify the atmospheric He content of a single
star like HD 15\,629 in only two Myr. 
(Another possible scenario for an increased He abundance, close binary
interaction, can most probably be discarded in the case of these three
particular objects, see Mason et al., \cite{mas98}).  
Turning the argument around, if rotation does not play a significant
role in the evolution of single massive stars, HD 15\,629 could become
rather similar to first HD\,14\,947 and then HD 210\,839 in one to two
Myr. Unfortunately, we cannot wait to confirm this hypothesis.

\begin{figure}
{\epsfig{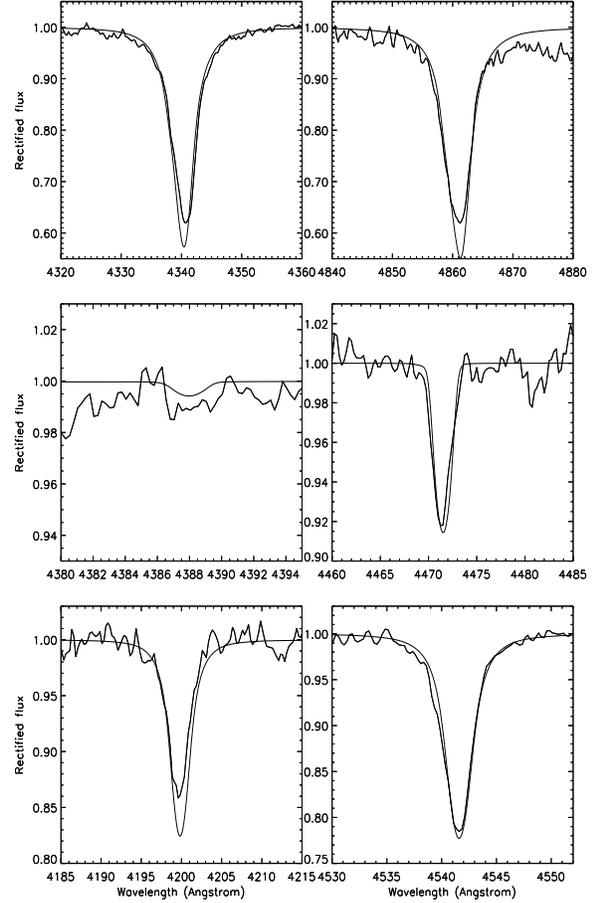}}
\caption[]{As Fig. 9, for HD 15\,629.}
\end{figure}

\subsection{HD 15\,570, O4 If$^+$} 
This star is hotter than the former one, and its Of signatures are
stronger (again, without being a transition object, see Conti et
al. \cite{con95}), indicating a lower gravity.  This is confirmed by
inspection of the Balmer series, especially the \Ha~ profile, which is
strong in emission. From 4600 to 4750 \AA~ we see a broad emission
feature with emission lines typical of Of stars, and both \SiIV~ lines
neighbouring \Hd~ are in emission, indicating a large luminosity,
which is confirmed. There is also absorption in
\NV\,$\lambda\lambda$4604, 4620.  It is by far the most luminous star
in the sample, 
and we were not able to fit the spectrum properly for this object.  The
extreme character of the features allows only a crude guess of the
stellar parameters, and we do not show any line fit. 
Apart from the projected rotational velocity of 105 km s$^{-1}$, we
can only say that the temperature is close to 50\,000 K and the
gravity should be of the order of \g\ $\sim$ 3.50 or even less.  With
the given luminosity, the evolutionary mass is very large, indicating
an initial mass in excess of 140 \Msun , which would make HD 15\,570
one of the most massive and luminous stars known in the Milky Way.
The spectroscopic mass is much lower, 
although we have to consider that the gravity has only been
``guessed'' so far (see next section).
There is a chance that the star is helium enhanced, and we
adopt \eps\ = 0.15.

\subsection{Cyg OB2 \#7, O3 If$^*$} 
This is a very hot star. \HeI~ $\lambda$4471 is only marginally seen,
and the emission in \NIV\,$\lambda$4058 and the absorptions in
\NV\,$\lambda\lambda$4604, 4620 are strong.  We again find the broad
emission feature with emission lines typical for Of stars, and both
\SiIV~ lines neighbouring \Hd~ in emission. In spite of our efforts,
we were unable to fit the star with plane--parallel models, the
required gravities being too low. As for HD 15\,570, we had to
extrapolate the plane--parallel parameters, and we do not show a line
fit. The final parameters we adopt are \teff\ = 51\,000 K, \g\ = 3.65,
\eps\ = 0.12 and \Vr\ = 105 km s$^{-1}$. We did not succeed in
calculating models in this range 
(\g~ had to be larger at least by 0.1), and spherical models including
mass-loss are obviously required.
This star is also very luminous and massive, with an initial mass in
excess of 110 \Msun , but with a large mass discrepancy.

\section{Analysis with spherical models with mass-loss}
\label{anas}
The difficulties in the analysis of some of the spectra discussed in the
previous section result from the neglect 
of sphericity and mass-loss effects. Thus, we decided to use the
program described by Santolaya--Rey et al. (\cite{sr97}) to account
for these effects.  Briefly, this program constructs a unified model
of the stellar photosphere and wind regions, with a number of standard
(stationarity, homogeneity) and non-standard 
assumptions and approximations (the most important of the latter
concerns the way the temperature structure is obtained,
which results in well approximated, but not forced, flux conservation
throughout the atmosphere and a constant temperature in the wind). It
solves the line-formation problem in an expanding atmosphere with
spherical geometry, and Stark broadening is included in the final formal
solution.  

The atomic models used are the H and He models used also in our
plane--parallel analysis, with minor changes to adapt them to the new
program.  Line-blocking has not been included so far. Its influence,
however, will be considered in the next section (at least on a
qualitative basis), and has affected the results decribed here.

We have adopted the following procedure. We begin with the parameters
given by the plane--parallel models (see Table 2).  The values of the
terminal velocities are taken from Puls et al.  (\cite{puls96}),
except for HD 15\,570 (from Lamers \& Leitherer \cite{ll93}) and for
Cyg OB 2 $\#$7 and HD\,5689, where we have adopted the \Vi\ --
spectral classification relation provided by Haser (\cite{has95}).
The last object has been assigned a \Vi~ corresponding to an O6 III
star, which is a compromise between the luminosity class V given by
Garmany \& Stencel (\cite{gs92}) and the low gravity obtained from our
analysis (see the discussion in the preceding section). Note, that
this compromise is reasonable, since the difference between the mean
\Vi~ of O6V and O6I stars is only 450 km s$^{-1}$.

Also needed are the values of the $\beta$ exponent in the
$\beta$--velocity law.  For HD 15\,558, HD 15\,629 and HD 210\,839 we
adopted the values given by Puls et al. (\cite{puls96}). For HD
14\,947, also analysed by Puls et al., we adopted $\beta$ = 1.15
instead of 1.0 as quoted by Puls et al., for reasons explained
later in the discussion of HD 14\,947. For Cyg OB2 $\#$7, HD 15\,570
and HD 5\,689 which were not analysed by Puls et al we used
$\beta$= 0.8, 1.0 and 1.0, respectively.

For each star we have constructed a small model grid by varying
gravity, temperature and mass--loss rate. We then tried to fit \Hg ,
\Ha~ and the ratio \HeII~ $\lambda$4200 to \HeI~ $\lambda$4471, which
is our new preferred temperature indicator, following the results
described in the next section, where we will show that this \HeII~
line is less sensitive to model details than \HeII~ $\lambda$4541. Of
course, the use of a new temperature indicator introduces also an
additional difference to the results from our plane--parallel
considerations.

All other parameters remain fixed at the beginning at their values
determined from the plane--parallel analysis. When necessary, the
helium abundance has been changed later. The radius merits additional
comments. Its value is derived from the emergent flux, which has been
obtained for a given set of parameters (those from the plane--parallel
analysis). Changing the parameters will change the derived radii, and
thus an iteration process might be required. We have recalculated the
radii of our objects with the new parameters, and have obtained an
average change of only 2$\%$ (the maximum change being of only
4.5$\%$, from 22 to 23 \Rsun). Thus we used the plane--parallel
values without further iteration.  
The finally adopted and derived parameters are summarized in
Table 3. For those stars which have been analysed by Puls et
al. (\cite{puls96}), we give their results in Table 4 for comparison.
The new position of the stars in the HR diagram is shown
in Fig. 13.

\begin{table*}
\label{parames}
\caption[ ]{Parameters determined for the programme stars starting
with the parameters from Table 2 and using
the spherical models with mass-loss. Temperatures are in thousands 
of Kelvin, velocities in km s$^{-1}$ and mass-losses in 
solar masses per year. MWM stands for modified wind momentum, and given
is \lMWM, with $\Mdot$ in solar masses per year, \Vi\, in
km s$^{-1}$ and $R$ in solar radii. \Ms , \Mev~ and \Mo~ are,  
respectively, the present spectroscopic and 
evolutionary masses, and the initial evolutionary mass,
in solar units. The last column indicates whether we formally have  a
mass discrepancy considering an error of 0.22 in log(\Ms ).}
\begin{flushleft}
\begin{tabular}{lccccccccccccc} 
\hline
Star & \teff~ & \g~ & \eps~ & \R~ & \lL~ & \Vi & $\beta$ & \lMp & \Ms~ &
\Mev~ & \Mo~ & MWM & M discr.?\\
\hline
Cyg OB2 \#7& 50.0 & 3.72 & 0.18 & 16.7 & 6.20 & 2900 & 0.80 & --4.95 & 53.0 &
104.5 & 107 & 29.93 & Yes\\
HD 15\,570 & 42.0 & 3.80 & 0.15 & 22.0 & 6.14 & 2600 & 1.05 & --4.75 & 112.6 &
79.6 & 89 & 30.14 & No\\
HD 15\,629 & 46.0 & 3.81 & 0.09 & 12.7 & 5.81 & 3000 & 1.00 & --6.13 & 37.5 &
61.4  & 63 & 28.70 & No\\
HD 15\,558 & 46.5 & 3.86 & 0.07 & 18.5 & 6.16 & 2800 & 0.75 & --5.40 & 89.5 &
91.7  & 97 & 29.48 & No\\
HD 14\,947 & 40.0 & 3.67 & 0.20 & 14.8 & 5.70 & 2400 & 1.15 & --5.25 & 36.8 &
47.5  & 49 & 29.52 & No\\
HD 210\,839& 37.0 & 3.55 & 0.25 & 18.9 & 5.78 & 2250 & 0.90 & --5.17 & 45.8 &
49.3  & 52 & 29.62 & No\\
HD 5\,689  & 37.0 & 3.57 & 0.25 &  7.7 & 5.00 & 2500 & 1.00 & --6.80 &  7.9 &
25.4  & 26 & 27.85 & Yes\\
\end{tabular}
\end{flushleft}
\end{table*}

\begin{table*}
\label{parames2}
\caption[ ]{Parameters as determined by Puls et al. (\protect\cite{puls96})
for some of the stars in Table 3. Units are as in Table 3.}
\begin{flushleft}
\begin{tabular}{lcccccccccc} 
\hline
Star & \teff~ & \g~ & \eps~ & \R~ & \lL~ & \Vi & $\beta$ & \lMp & \lMWM \\
\hline
HD 15\,629 & 47.0 & 3.90 & 0.07 & 14.2 & 5.95 & 3000 & 1.00 & --6.12 & 28.73\\
HD 15\,558 & 48.0 & 3.85 & 0.07 & 21.8 & 6.36 & 2800 & 0.75 & --5.14 & 29.78\\
HD 14\,947 & 43.5 & 3.50 & 0.15 & 16.1 & 5.93 & 2350 & 1.00 & --5.12 & 29.65\\
HD 210\,839& 38.0 & 3.65 & 0.09 & 19.0 & 5.83 & 2250 & 0.90 & --5.28 & 29.52\\
\end{tabular}
\end{flushleft}
\end{table*}

\begin{figure}
{\psfig{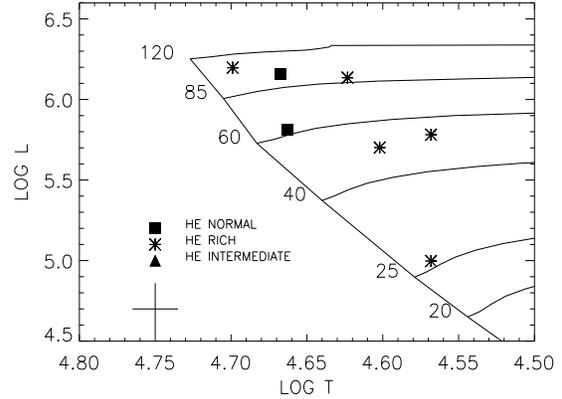}}
\caption[]{The programme stars on the Hertzsprung--Russell diagram
after our analysis with spherical models with mass-loss. Compare
the new diagram with Fig. 7. Details are as in Fig. 7.}
\end{figure}

\subsection{HD 5\,689, \bf O6 V}
We can see the line fit to this star in Fig. 14, for a model with
\teff\ = 37\,000 K, \g\ = 3.45 (which increases up to 3.57 due to the
centrifugal correction) and \lMp\ = --6.80.  The temperature is lower
than in the plane--parallel case, in part due to the change of our
temperature criterion. The gravity, however, remains the same (which
actually implies an increase with respect to plane--parallel models,
since a lower temperature usually demands a lower gravity to fit a given
line). Thus, the large mass discrepancy found above (see Table 2)
cannot be reduced by the spherical models, since it originates
from the low value of the radius, rather than from the low gravity.
Neither can we decide whether the star has a larger radius in parallel
with a higher mass--loss rate, since the \Ha\, wind-emission depends
only on the ratio (\.M /$R^{3/2}$), whereas its photospheric component is
independent of both.
Assuming an absolute magnitude of a typical
O6 III star ($M_V$ = --5.78 mag, see Table 6 of Vacca et al.
\cite{vacc96}) results in a radius of 15.9 \Rsun , but does not change
the fit quality (it affects only slightly the fit of \HeI~
$\lambda$4471). Of course, mass, luminosity
and mass-loss rate 
would be affected and we would obtain a spectroscopic mass of 30.0
\Msun , an evolutionary mass of 41.9 \Msun~ (thus reducing the mass
discrepancy to 30$\%$, a typical value), a luminosity of 5.6,
closer to the other objects (see also the discussion in
Sect.~\ref{discu}) and a logarithmic mass-loss rate of -6.17. 
However, this possibility would also imply that
the wind in HD\,5689 is much less efficient than in HD\,210\,839, as
both stars would have roughly similar parameters, except for $\Mdot$.
Unfortunately, without knowing the distance to HD\,5689 more
accurately, we cannot derive stronger conclusions. We also see that
\HeII~ $\lambda$4541 does not fit completely, and in particular the
fit of \HeII~ $\lambda$4686 is poor. This line always shows a poor
fit, with an observed absorption that is much stronger than predicted,
especially in the blue wing. As explained in the following section,
this seems to be related to lack of line-blocking in our models, and
\HeII~ $\lambda$4541 seems also to be affected, although to a lesser
extent.

\begin{figure}
{\psfig{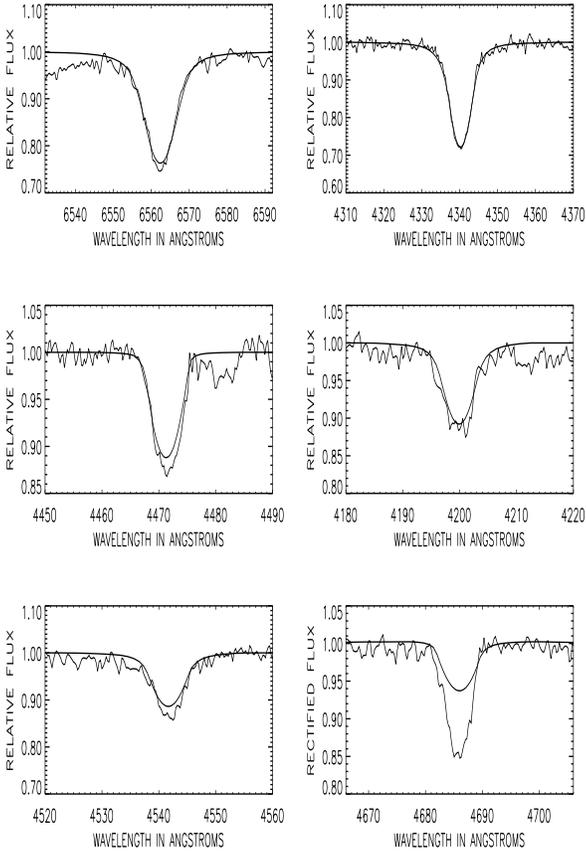}}
\caption[]{Fit to the H and He lines in HD 5\,689 using a
spherical model with the parameters given in Table 3. 
From left to right and top to bottom we show \Hg, \Ha, \HeI~4471,
\HeII~ 4200, \HeII~ 4541 and \HeII~ 4686. Wavelengths 
(along the abscissa) are given in \AA . The ordinates give the relative fluxes.}
\end{figure}

\subsection{HD 210\,839, O6 I(n)fp}
The line fit of this star, shown in Fig. 15, reveals two
problems. First, the form of \Hg\, suggest that the adopted rotational
velocity is too large. A value of 200 km s$^{-1}$ results in a much
better fit\footnote{a misprint in Table 8 of Puls et
al. (\cite{puls96}) gives a velocity of 100 km s$^{-1}$ for this star,
whereas 200 km s$^{-1}$ was the value adopted in that work.  and
would agree with previous determinations (Conti \& Ebbets,
\cite{coneb77}, Penny, \cite{penny96}, Howarth et al., \cite{how97}).
A change in the rotational velocity would not change the adopted
parameters, only the quality of the line fit. We show the fit with the
value adopted for the plane--parallel analysis, although a better fit
is possible with the lower \Vr\, value.}

The second problem is that we were unable to fit the P Cygni form 
of the \Ha--profile. 
As was pointed out previously, $\lambda$ Cep is a possible non-radial
pulsator (Fullerton et al. \cite{fgb96}, de Jong et al. \cite{dJo99}),
which might induce deviations from homogeneity (by exciting the
line-driven wind instability already in the lower wind part, see
Feldmeier et al. \cite{feld97}),
and the large rotation rate might have an additional influence on the
wind structure and the resulting profile (cf. Owocki et al. \cite{owo98}
and references therein; Petrenz \& Puls \cite{petpuls96}).
We decided to concentrate on the red wing of \Ha , since the
theoretical simulations result in a blue wing affected by an extra
\HeII~ absorption, which is inadequately described in our present
models (see next section), and any wind variability becomes much more
visible in the blue wing, compared to the red one.
Again, the predicted emission in \HeII~ $\lambda$4686 is much stronger
than observed; also, the temperature is lower than in the
plane--parallel case.  However, the gravity is larger, and we do not
find a mass discrepancy for this star.

$\lambda$ Cep has been also analysed by Puls et al. (\cite{puls96}),
who adopted slightly different parameters. The main difference is the
helium abundance (\eps\ = 0.09 instead of 0.25), which has a small
influence on the gravity, however produces a larger wind momentum with
a lower luminosity in our present results. (The value for \eps\, quoted
by Puls et al. was not derived from a consistent spectral analysis,
however taken from the literature). It is interesting to note that the
high abundance favoured by our findings is in agreement with Blaauw's
suggestion that all runaway stars are He enriched (Blaauw,
\cite{bla93}). Note also that the possible runaway nature of
HD\,5\,689 fits within this scenario.

Finally, we point out that our mass--loss rate coincides well with
other determinations from \Ha\ by Lamers \&
Leitherer \cite{ll93}, but is larger than that derived from radio
fluxes (Lamers \& Leitherer \cite{ll93}) by a factor of 3.

\begin{figure}
{\psfig{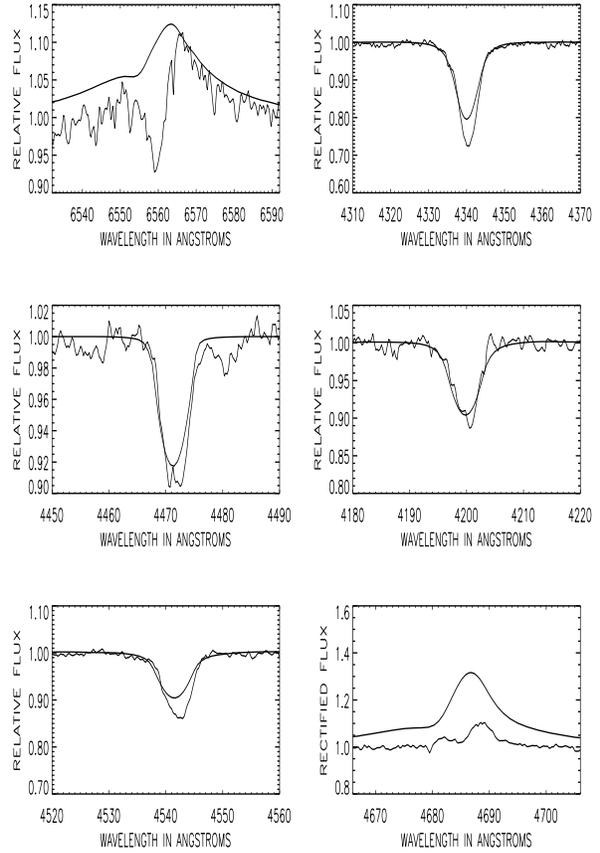}}
\caption[]{As Fig. 14, but for HD 210\,839. The fit of \Hg~
improves considerably if the adopted rotational velocity is lowered
from 250 to 200 km s$^{-1}$.}
\end{figure}
 
\subsection{HD 14\,947, O5 If$^{+}$}
The analysis of HD 14\,947 has been hindered by an inconsistency in
the radial velocities derived from the blue and the red spectrum
(the spectra were taken in September 1991 and August 1992, see
Table 1). While in the blue spectrum there are several
metal lines to derive the radial velocity correction, in the
red spectrum we only have three lines of \HeII~ and they give
a result that is incompatible with that of the blue metal lines.
Mason et al. (\cite{mas98}) list this star as having a constant radial
velocity and thus we don't have a clear explanation for the 
radial velocity change. Therefore,
the radial velocity correction for this star is particularly
inaccurate. We have adopted the correction indicated by the blue 
metal lines, accounting for the different rest frames as a function
of observation date.

The line fit of \Hg~ is comparatively poor for this star (see
Fig. 16). This reflects a problem that we have found for the first
time, an inconsistency between the fit for \Hg~ and \Ha . The fit for
\Hg~ would need a mass-loss rate which is only half that of \Ha. This
inconsistency is only weakly dependent on the gravity.

The new temperature derived for HD 14\,947 is much lower than that
obtained from plane--parallel models.  This is due to sphericity
together with the new criterion of using \HeII~$\lambda$ 4200.  Again,
in agreement with the large mass--loss rate,
\HeII~$\lambda\lambda$4541, 4686 are poorly fitted, (see discussion in
the next section).

The spectroscopic mass derived is clearly lower than the evolutionary
one, although they agree within the errors. The analysis by Puls et
al.  (\cite{puls96}) gave slightly different stellar parameters, the
most important difference being the now lower temperature implying a
lower luminosity by 0.23 dex.

There is an upper limit of 10$^{-4.75}$ \Msun/yr for the mass--loss
rate of HD\,14\,947 derived from radio fluxes (Lamers \& Leitherer
\cite{ll93}). The value we find here is in agreement with this upper
limit.

\begin{figure}
{\psfig{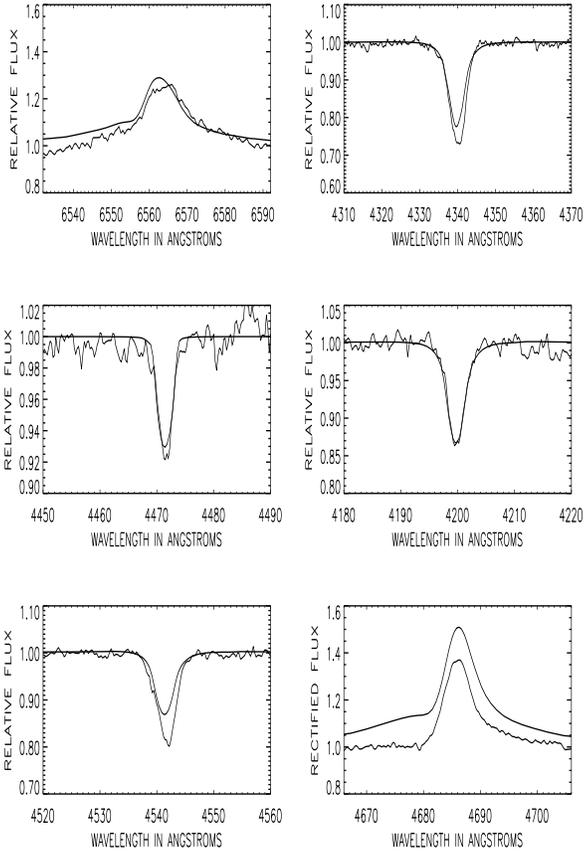}}
\caption[]{The line fit to HD 14\,947 using spherical models
with mass-loss. We see the comparatively poor
fit for \Hg\ in the core, which can be largely improved by
reducing the mass--loss rate by a factor of two.}
\end{figure}

\subsection{HD 15\,558, O5 III(f)}
The line fit to HD 15\,558 obtained for the parameters given in Table 3
can be seen in Fig. 17. The mass--loss rate is rather large, in
agreement with the high luminosity, especially when compared with HD
15\,629, a star that could be considered similar at first inspection
since both \Ha~ profiles are in absorption. However, the value we
obtain here for $\Mdot$ is lower than that obtained by Puls et al.
(\cite{puls96}) due to the change in the stellar parameters.  The
gravity is largely increased with respect to the value derived with
plane--parallel, hydrostatic models (by 0.15 dex), and in fact the
derived spectroscopic mass is nearly equal to the evolutionary one,
thus making HD 15\,558 one of the few cases for which we do not find
any mass discrepancy. The generally good agreement in the line fit to
HD 15\,558 is again broken by \HeII~$\lambda$4686.

\begin{figure}
{\psfig{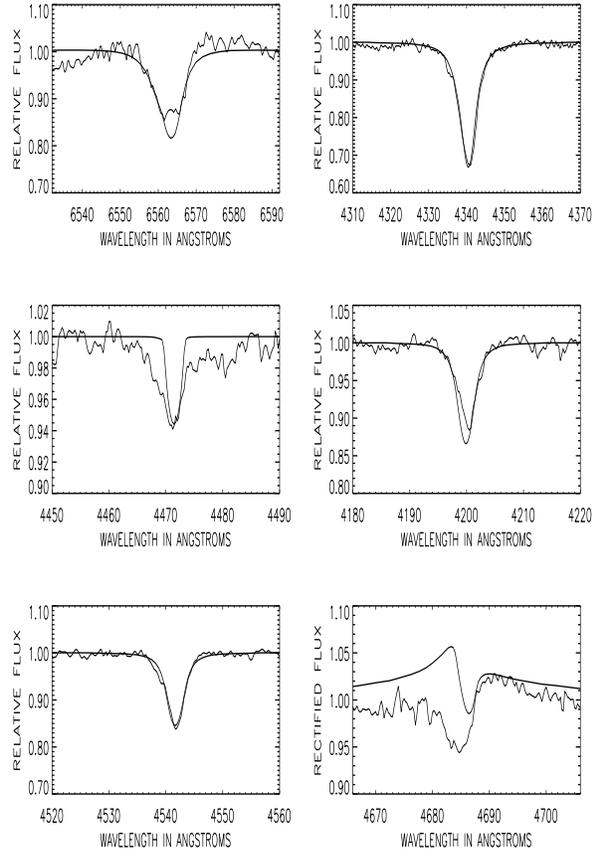}}
\caption[]{As Fig. 14, but now showing the line fit to HD 15\,558.}
\end{figure}

\subsection{HD 15\,629, O5 V((f))}
The line fit of HD 15\,629 is obtained for the parameters given in
Table 3 and is shown in Fig. 18.  We see that the mass--loss rate is
relatively low and agrees very well with the value given by Puls et
al. (\cite{puls96}).  The gravity we obtain is similar to that in the
plane--parallel case, and thus the new mass is again lower than the
evolutionary one, although there is formal agreement considering the
error bars. The worst fit corresponds to \HeII~$\lambda$4686.  A
comparison with Puls et al. (\cite{puls96}) shows that both sets of
values are compatible, although the general trends (lower
temperatures, radii and luminosities in our case) remain.

\begin{figure}
{\psfig{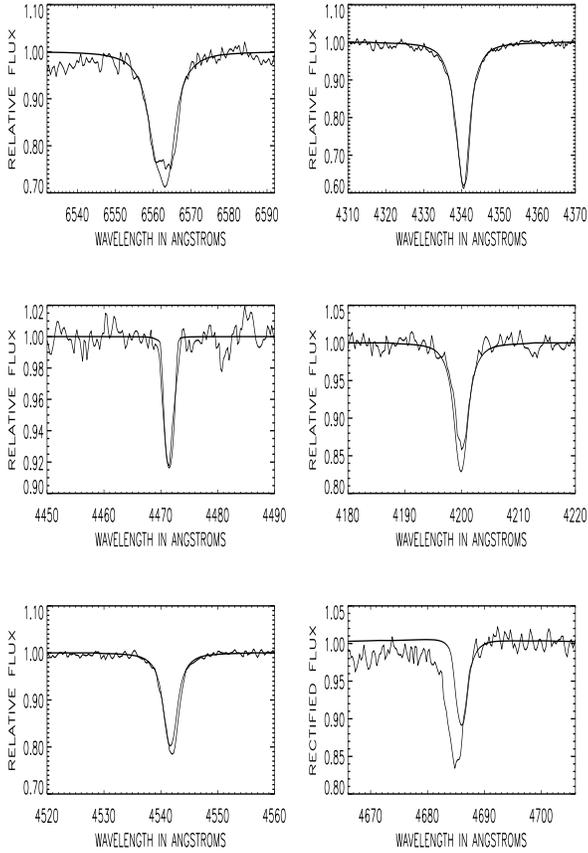}}
\caption[]{As Fig. 14, for HD 15\,629.}
\end{figure}

\subsection{HD 15\,570, O4 If$^{+}$}
This is an extreme object, and the adopted parameters are actually a
compromise, reflected in very large uncertainties.  For this star we
derive \teff\ = 42\,000 K, \g\ = 3.80, \eps\ = 0.15, \lMp\ = -4.75
(see Table 3 for the rest of the parameters).  The very high mass--loss
rate results in an extreme insensitivity of the wings of \Hg~ to
gravity variations, as can be seen in Fig. 19, where we have plotted
the profiles for \g\ = 4.05 and \g\ = 3.55. These can be considered as
limits beyond which the fits begin to become poorer than for the range
[3.55,4.05]. Thus we have adopted \g\ = 3.80$\pm$0.25. However, it
should be borne in mind that even such a large error is optimistic, as
the uncertainty of the radial velocity correction has not been
included. Also the He profiles are largely insensitive to \g~
variations, but they allow us to restrict \teff~ and \eps .
 
\begin{figure}
{\psfig{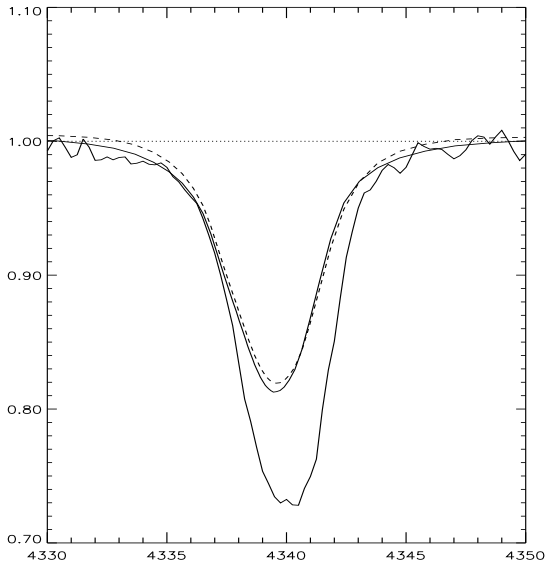}}
\caption[]{The line fit to \Hg~ in HD 15\,570 with \g\ = 4.05 (solid
line) and \g\ = 3.55 (dashed line). All other parameters are as in
Table 3.  The abscissa gives wavelength in \AA\, and the ordinate
relative fluxes.}
\label{compgrav}
\end{figure}

The line fit to this star is shown in Fig. 20.  A detailed fit of the
strong \Ha~ emission of HD 15\,570 was not possible, since for the
given parameters the simulations always showed a double peak due to
the extra emission coming from the \HeII~ blend (see the discussion in
the following section) not seen in the observations. Thus the adopted
criterion was to fit the red wing of \Ha . Again, we found the same
problem encountered for HD 14\,947.  The best fit of \Hg~ gives a
mass--loss rate different from the best fit of \Ha , without any
possibility to fit both simultaneously, with a discrepancy of roughly
a factor of two.
\Ha~ indicates a value of about \lMp\ = --4.75 and \Hg~ one around
\lMp\ = --5.0, but this low value of \lMp~ would also imply that the
gravity is lower than in the case of higher mass--loss rate.  This is
shown in Fig. 21, where we see the fit for \Hg~ at \lMp\ = --4.75 and
--5.0 at \g\ = 3.80. Comparing with Fig. 19 we see that the effect of
lowering the mass--loss rate is larger than any change of gravity at
\lMp\ = --4.75. At the lower \lMp~ of --5.0 we would obtain a lower
and more constrained gravity, but it is impossible to obtain even a
moderate fit to \Ha . In any case, the mass--loss rate is very large,
although we should point out again that our mass--loss rate coincides
well with other \Ha\, determinations (Puls et al. \cite{puls96};
Lamers \& Leitherer \cite{ll93}) but it is larger than that derived from
radio fluxes (Lamers \& Leitherer \cite{ll93}) by a factor of 4 if we
adopt the value from the fit of \Ha , and by a factor of 2 if we take
the one more consistent with \Hg .

The temperature is now much lower than in the plane--parallel case, as
for HD 14\,947, again due to the combined effects of sphericity and
the new criterion for the He ionization equilibrium.  In agreement
with this, the fit to \HeII~$\lambda\lambda$4541, 4686 is poor.  The
formal change in \g~ is large with respect to the plane--parallel
values (0.20 dex) and now HD 15\,570 has also the largest
spectroscopic mass, even formally exceeding 100 \Msun . Little can be
said about the actual mass and the mass discrepancy. In Table 3 we see
that the spectroscopic mass is 30$\%$ larger than the evolutionary
one, but we did not attemp to bring both into agreement (which could
be possible adopting \g\ = 3.65), since for HD 15\,570 gravity and
mass are only formal values in the centre of a large uncertainty area,
and have to be regarded as rather inaccurate.  It is clear however that HD
15\,570 is one of the most extreme O stars in the Milky Way, and
probably one of the most massive stars.

\begin{figure}
{\psfig{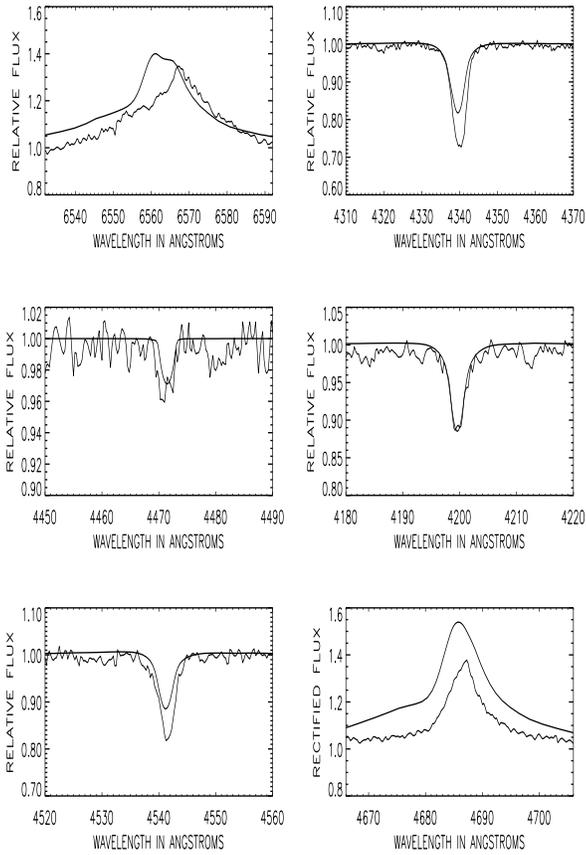}}
\caption[]{The line fit for HD 15\,570. The fit shown corresponds to the
model fitting the red wing of \Ha~(see text).}
\end{figure}

\begin{figure}
{\psfig{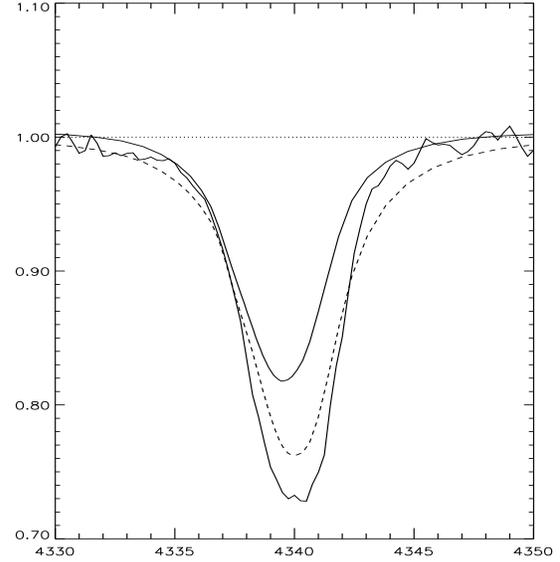}}
\caption[]{The line fit to \Hg~ in HD 15\,570 with \lMp\ = --4.75
(solid line) and \lMp\ = --5.00 (dashed line) at \g\ = 3.80.  All
other parameters are as in Table 3. Comparing with Fig.~\ref{compgrav}
we see that for this strong wind, changing the mass--loss rate has a
larger effect than changing the gravity. The abscissa gives wavelength
in \AA\, and the ordinate relative fluxes.}
\end{figure}

\subsection{Cyg OB2 $\#$7, O3 If$^{*}$} 
The line fit for this star, displayed in Fig. 22, shows similar
problems to those of HD 14\,947 and HD 15\,570 discussed before.  The
mass--loss rate is quite large, again exceeding 10$^{-5}$ \Msun /yr,
and the parameters point to a very massive star. The large mass--loss
rate results in a change of \g~ with respect to the plane--parallel
hydrostatic models, which is relatively large (0.15).  However, the
mass discrepancy is still present, since the spectroscopic mass is
still a factor of 2 lower than the derived evolutionary mass.  Another
important change with respect to plane--parallel models is the new
helium abundance, now \eps= 0.18. It is interesting to note that
Herrero et al. (\cite{h99}) could not find evidence of any helium
discrepancy in their analyses of other, less extreme Cyg OB2 stars.

\begin{figure}
{\psfig{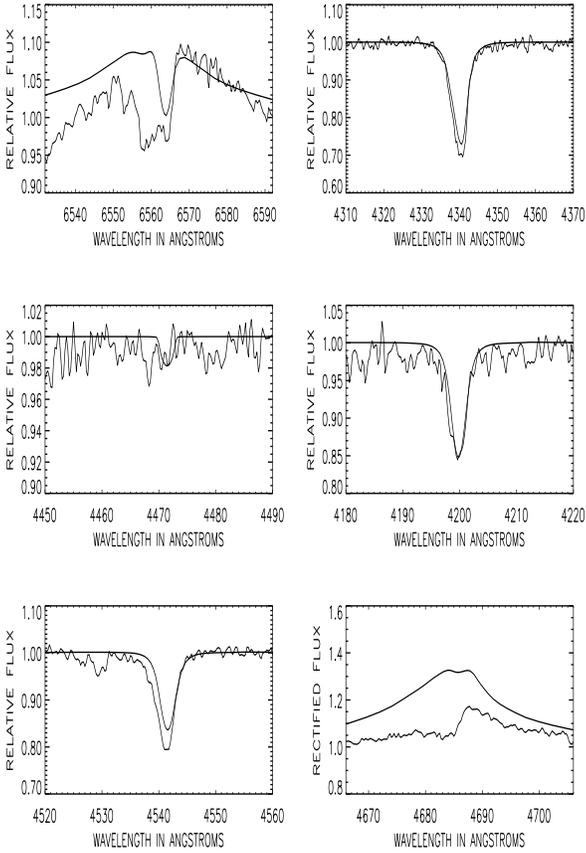}}
\caption[]{The line fit for Cyg OB2 $\#7$ using a
spherical model with the parameters given in Table 3.}
\end{figure}
 
\section{Additional considerations}
\label{numef}
\subsection{The H$_\gamma$ problem}
As we have seen in the foregoing analysis, 
for three of our objects with dense winds the wind emission seen in
H$_\gamma$ is inconsistent with the mass--loss rate derived from
H$_\alpha$. It may be questioned why this obvious problem has not
already been discovered by Puls et al. (1996) who found no such
discrepancy and derived--- for the object common to both investigations
(HD~14\,947\footnote{Note, that HD~15\,570 has not been analysed by
Puls et al. by means of detailed profile fits. Instead, only
previously published values for equivalent widths have been used, so
that any comparison is obsolete.}) ---a lower values for \g~ than
follows from our analysis.

This difference has the following origin. For the problematic objects,
the Doppler {\it core} of \Hg\, becomes optically thick in the wind.
Thus, it depends only on the wind conditions (mainly in the transonic
region) and is independent of any underlying photospheric radiation.
Consequently, it can be correctly fitted by the approach described in
Puls et al. and would compare to our results.

On the other hand, the {\it wings} of \Hg~ depend on the illuminating
radiation from below the transonic region. In the concept applied by
Puls et al., this radiation was taken from photospheric profiles
calculated on the basis of {\it hydrostatic} models, and the wind
contamination was correctly accounted for. The fitted \g~ value was
finally corrected for the difference between hydrodynamic and
hydrostatic atmospheres, accounting for the different formation depths
in an approximate and {\it global} way, where the correction turned
out to be only moderate in the parameter space considered.

This procedure, however, is only justified if the major part of the
wings is actually formed in some pseudo-hydrostatic environment, which
is the case if the wind densities are not too high. For increasing
wind density, however, the difference between hydrostatic and
hydrodynamic stratification (e.g., Puls et al., Fig.~16) becomes
increasingly larger, and the point where the transition between both
regimes occurs shifts to correspondingly larger (mean) optical
depths. Thus, for higher wind densities only the far wings are formed
in a purely hydrostatic environment, whereas the inner wings are
severely affected by the wind conditions, which display a lower
density at given optical depth. Consequently, the use of hydrostatic
profiles and the global correction applied by Puls et al. will
inevitably fail under those conditions. Moreover, the derived values
for \g~ will be too small compared to ``reality'', since, for a given
\g, the hydrostatic densities are always larger than those in the wind
regime. As a result, the deviations between the profiles calculated by
Puls et al. and ours should become largest just outside the Doppler
core, and vanish at the extreme wings.  Thus, it is not too surprising
that the results of our {\it consistent} description deviate from the
results given by Puls et al. with {\it respect to gravity} (derived
from the wings), whereas the \Hg\, Doppler cores would be consistent
with the mass--loss rate derived from \Ha, 
however are unfortunately not visible due to rotational broadening.

We like to point out that for a (small) number of objects with denser
winds the determination of the stellar mass becomes completely
impossible, since in those cases even the continuum is formed in the
wind and the reaction of {\it any} profile on pressure scale height
becomes impossible.

\begin{figure}[t]
{\psfig{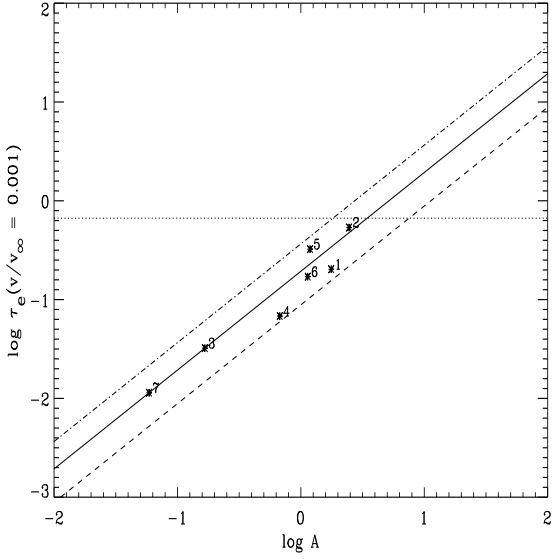}}
\caption[]{Electron scattering optical depth at transition velocity $v/\vinf =
0.001$ as function of $A$ (cf., Eq.~\ref{defa}) and $\beta$ = 1.3
(dashed-dotted), 1.0 (fully drawn) and 0.7 (dashed). The positions of our
programme stars (according to the derived wind parameters) have been indicated
by asterisks.}
\label{plottaue}
\end{figure}

In order to check under which conditions this problem will arise, we
have calculated the (minimum) continuum optical depth given by
electron scattering as function of wind parameters:

\beq
\taue = \int_r^{\inf} \se \rho(r) \dd r, \,\,
\se \approx 0.4 \,\mbox{cm$^2$/ g} \, \frac{1 + \Ihe Y}{1+4Y},
\eeq
with $\se$ the electron scattering mass absorption coefficient, $Y$
the helium abundance (by number) with respect to hydrogen and $\Ihe$
the number of free electrons provided per He atom. By means of the
equation of continuity and for the $\beta$ velocity law which is used
also in our model simulations,

\beq
v(r) =  \vinf \bigl(1 - \frac{b \Rstar}{r} \bigr)^\beta, \,\, 
b=1-(v_o/\vinf)^{1/\beta}
\eeq
(where $v_o \approx 0.1 v_{\rm sound}$ is the velocity at the transition
point between pseudo-hydrostatic and wind region, cf Santolaya--Rey et al.,
\cite{sr97}, their Sect.~2.2), we have 

\beq
\taue(v) = \frac{\Mdot \se}{4 \pi \Rstar \vinf} \frac{1}{b \beta}
\int^1_{v/\vinf} v^{\prime\,({1 \over \beta} - 2)} \dd v^{\prime},
\eeq
where $v^\prime= v({\rm r})/\vinf$.
Inserting typical parameters and denoting by $A$ the optical depth-like
quantity

\beq
\label{defa}
A = \frac{\Mdot}{10^{-6} {\rm \Msuna/yr}} \, \frac{10\Rsuna}{\Rstar}\,
\frac{1000\,{\rm km/s}}{\vinf}\, \frac{1+\Ihe Y}{1+4Y},
\eeq
the electron scattering optical depth is finally given by

\beqa
\taue(v) & \approx & 0.028\,A \,(-\ln\frac{v}{\vinf}) \,\,\,\,{\rm for}\,\beta =1
\nonumber \\
&\approx& 0.028 \, \frac{A}{1-\beta}\, 
\Bigl(1 -(\frac{v}{\vinf})^{\frac{1-\beta}{\beta}} \Bigr),\,\,\beta \ne 1
\eeqa
(approximating $b$ by unity). In Fig.~\ref{plottaue} we have plotted
this quantity, evaluated at $v/\vinf = 0.001$ (which compares roughly
with the velocity at the transition point for hot stars) as a function
of $A$ and $\beta$. The dotted line gives the optical depth of 2/3,
and for all objects with $\taue(A) > 2/3$ the continuum is definitely
formed in the wind. We have indicated the position of our sample stars
by asterisks, where the number refers to Table~\ref{obs}. Obviously,
object no. 2 ( = HD 15\,570) lies just at the border line, and thus a
gravity determination is almost impossible (cf., Fig. 19 for the
influence of gravity on \Hg). For the objects nos. 1 (Cyg OB2 $\#$7),
5 (HD\,14\,947) and 6 (HD\,210\,839) at least the inner wings of \Hg\,
are severely affected by the wind, and thus result in a larger gravity
than found by Puls et al.  (if all other parameters remain the same).
For the remaining three stars, the continuum is formed solely in the
hydrostatic part, and the derived numbers should coincide with the
approximate method, as is actually the case.

\subsection{Influence of He{\sc ii} resonance lines}

Those readers in particular who are not well familiarized with
spectral analyses of hot stars may question why we did not take the
most prominent line of He{\sc ii} in the optical part of the spectrum,
namely He{\sc ii}~4686 ($n = 3 \rarrow 4$), into consideration so far
(although we have plotted it for all models and the discrepancy is
obvious). This question is completely legitimate, since the upper
level of this transition is just the lower one of our strategic lines
\HeII\, $\lambda\lambda$ 4200, 4541, and should be reproduced with a
similar degree of precision if our models were reliable.

However, it is well known that this line (if formed in the wind) is
extremely difficult to fit, and, to our knowledge, has never been used
in any kind of NLTE-analysis of luminous O-stars. Usually, if one
compares the predicted profiles to observations, the synthetic line
turns out to be too strong in emission, even if all other lines
including \Ha\, do perfectly fit.

This rather unsatisfactory behaviour, which is normally by-passed by
simply excluding He{\sc ii} 4686 from the line list, relates to the
extreme sensitivity of the participating levels on the treatment of
the He~{\sc ii} resonance lines and their sensitivity to line-blocking
(for a discussion concerning this problem of the formation of He~{\sc
i}-lines, cf., Santolaya--Rey et al. \cite{sr97}; see also the related
discussion concerning the ionization structure of WRs by Schmutz
\cite{Schmutz97}). In standard simulations for wind conditions as
described here, where line-blocking effects are excluded, the
dominating background ``opacity'' below 303 \AA\, (referring to He{\sc
ii} Ly$_{\alpha}$) is Thomson {\it scattering}, leading to extremely
enhanced radiation temperatures at the resonance-line'
frequencies. Compared, for example, to a detailed balance situation
(see below), the ground-state becomes depopulated, which in
consequence (and in connection with the increased escape-probabilities
due to the velocity field) prohibits He~{\sc iii} from recombining and
gives rise to much weaker absorption edges at 229 \AA, compared to
plane--parallel simulations (cf., Gabler et al. \cite{Gableretal89},
especially Fig.~A2).

Moreover, since the radiation temperature is increasing towards higher
frequencies (due to the decreasing bf-opacity at lowest photospheric
levels), the NLTE departure coefficients are larger for higher levels
than for lower ones. Thus, in addition to the wind emission by
geometrical effects, the lines between excited levels (predominately
He~{\sc ii} 4686) are contaminated by a strong source function
$\propto b_{\rm upper}/b_{\rm lower} > 1$, which leads to a much
stronger total emission than would be the case if the resonance lines
were of less importance. Actually, a pilot investigation by Sellmaier
(\cite{Sell96}) for the case of $\zeta$ Pup has shown that the
emission of He~{\sc ii} 4686 could be significantly reduced if
line-blocking was accounted for correctly.

In order to investigate in how far the above effects are of influence
for our analysis (especially for the strengths of He~{\sc ii} 4200 and
4541, respectively), we have run a number of simulations with
different treatments of the He~{\sc ii} resonance lines, for the
example of our final model for HD~14947, where the majority of lines
is formed in the wind. To check our hypothesis that the dominating
effect leading to erroneous results follows from the increased pumping
by resonance lines, two principally different approaches were
considered, which should give similar results if the hypothesis were
correct.

On the one hand, we set all He{\sc ii} resonance lines into detailed
balance. Alternatively, we simulated an additional $\rho$-square
dependent background opacity in the decisive frequency range 227 \AA\
$< \lambda <$ 400 \AA, defined by

\beq
\label{defkappa}
\chi_{\nu}^{\rm add} = \kappa \xne \cdot \se \rho \cdot f_{\nu},
\eeq
with different values of $\kappa$ between $3 \cdot 10^{-14}$ and $1 \cdot
10^{-16}$.  The frequential dependence $f_{\nu}$ was assumed to be either
increasing or decreasing,

\beqa
f_{\nu}^1 & = & 1 - q_{\nu} \nonumber \\
f_{\nu}^2 & = & 0.5 (1 + q_{\nu}) \nonumber \\
q_{\nu} & = & \frac{1/\lambda - 1/227}{1/400 - 1/227},
\eeqa
and the appropriate emission component has been set to Planck.  The
value of $\kappa$ corresponds to the inverse of the electron density
at that point where the additional opacity reaches the same value as
the electron scattering opacity. From the numbers given above, it is
obvious that our choice is rather low compared to what might be
expected in reality. For our final discussion, we have selected four
models with parameters given in Table~\ref{modres}.

\begin{table} 
\caption[]{Different approximations for treatment of He{\sc ii} resonance
lines: The value of $\kappa$ corresponds to the definition in Eq.~
(\ref{defkappa}), and the line styles are the same as in Figs.~\ref{plot_trad}
and \ref{plot_1640}.}
\label{modres}
\begin{center}
\begin{tabular}{lccrr}
\hline \hline
no.  & approx. & $\kappa$ & $f_{\nu}$ & line style \\ 
\hline
1 & ``standard'' & -- & -- & dotted \\
2 & detailed balance & -- & -- & dashed \\
3 & blocked flux & $3 \cdot 10^{-14} $ & $f_{\nu}^2$ & fully drawn \\
4 & blocked flux & $3 \cdot 10^{-16} $ & $f_{\nu}^1$ & dashed-dotted \\
\hline
\end{tabular} 
\end{center} 
\end{table}

\begin{figure}[t]
\resizebox{\hsize}{!}
   {\includegraphics{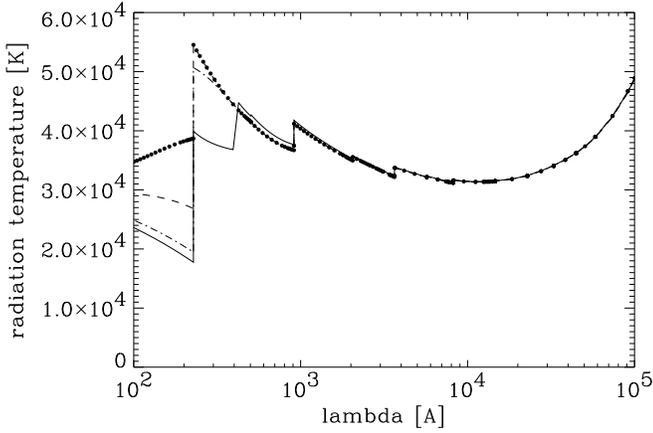}}
\caption[]{Radiation temperatures of emergent fluxes for
the model of HD~14947, with different treatment of He{\sc ii} resonance lines.
Line styles as defined in Table~\ref{modres}.}
\label{plot_trad}
\end{figure}

Fig.\ref{plot_trad} verifies the expected behaviour for the He{\sc
ii} ground-state. Both for the models with detailed balance as well as
with simulated background opacities, He{\sc iii} begins to recombine
in the outer atmosphere as long as the strong upward rates present in
model 1 (dotted) are no longer active, so that a significantly
enhanced ionization edge develops. Only for the model with the lowest
value of $\kappa = 1 \cdot 10^{-16}$ (not displayed), the influence of
the background opacity becomes so weak that the model remains ionized
throughout the wind.

\begin{figure}
{\psfig{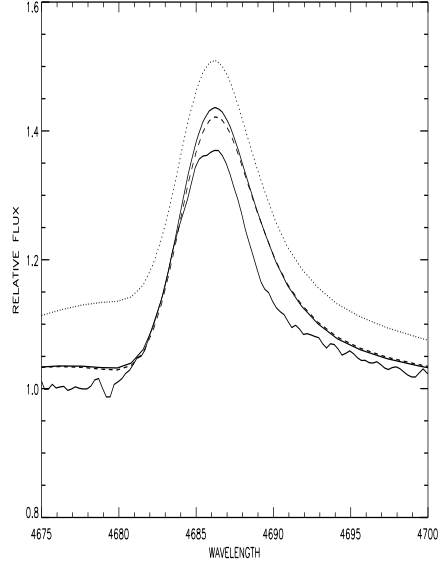}}
\caption[]{Comparison of the \HeII\,4686 profile observed in HD\,14\,947
with those obtained in our simulation of line--blocking (full drawn),
of detailed balance for \HeII\, resonance transitions (dashed)
and in our standard model (dotted).}
\label{heii}
\end{figure}

Besides the reaction of the ground-state, also the expected behaviour
of the excited states (reversal of population) takes place as long as
the radiation temperature does not significantly increase towards
higher frequencies (model 3) or the resonance lines are not active
(model 2). A comparison between the corresponding He{\sc ii} 4686
profiles and the observations (Fig.~\ref{heii}) shows that our
simulations are in almost perfect agreement, whereas our standard
model displays much too much emission. It is interesting to note here
a remark given by our anonymous referee. Henrichs (\cite{hen91})
reports that the equivalent width of \HeII\,$\lambda$4686 in
$\lambda$Cep varies in concert with the high--velocity edge of the
\CIV\,$\lambda$1550 line, suggesting that both variations share a
common origin that could rely on the behaviour of the background
opacity or the resonance lines of \HeII, as studied here.

A final check on how far our models compare to reality allows the
comparison with the UV-line He{\sc ii} 1640 ($n = 2 \rarrow 3$), since
the lower state of this transition behaves differently from the other
excited levels. In those cases where the resonance lines are no longer
active (detailed balance) or are of negligible importance (background
opacities) and the He{\sc ii} Lyman edge becomes optically thick
throughout the atmosphere, this state becomes the effective ground
state of the ion\footnote{This situation corresponds to the behaviour
of \Ha~ in A-type supergiants.}. Thus, its population is predominantly
controlled by the photoionization balance at 911 \AA\, coinciding with
the hydrogen Lyman edge. Since this edge is optically thin, the 2nd
level becomes strongly overpopulated because of the diluted radiation
field, i.e., the absorption should be larger than for the standard
model. A comparison with the observed {\it IUE} profile\footnote{SWP
10724, kindly provided in reduced form by I.D.~Howarth and
R.K.~Prinja.}  shows that our models are on the right track, in
contrast to the standard model which predicts too little
absorption. Only at higher velocities, \ie in the outermost wind, are
they too strong compared to observations, whereas for the inner wind,
which is the decisive part concerning our analysis, they are in
perfect agreement. A comparison with model 4 shows that some
fine-tuning might improve even the situation at larger velocities. (We
note that in order to fit the position of the emission peak, we had to
apply an artificial velocity dispersion of roughly 100 km/s,
consistent with the values found from the analysis of UV resonance
lines.)

\begin{figure}[t]
\resizebox{\hsize}{!}
   {\includegraphics{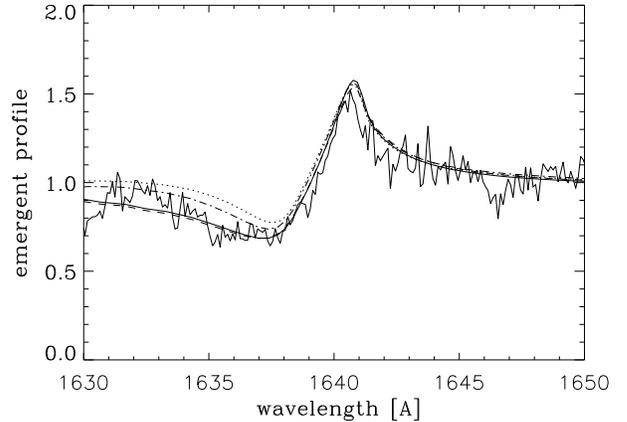}}
\caption[]{{\it IUE} spectrum of HD~14947\footnotemark[5], He{\sc ii}$\lambda$1640,
compared to profiles generated from models with different treatment of
He{\sc ii} resonance lines. Observations shifted by 90 km/s to the blue,
velocity dispersion $v_{\rm turb}$ = 100~km/s applied to account for the
red-shifted emission peak (see text). Line styles as defined in 
Table~\ref{modres}.}
\label{plot_1640}
\end{figure}

In conclusion, we found that suppressing the large upward rates from
resonance lines present in our standard model gives rise to a
different population of ground and excited levels. The exact mechanism
for this suppression, however, seems to be irrelevant to the results,
and the synthesized profiles of previously problematic lines compare
well with the observations.

We are now able to check the consequences of the manipulations
outlined with respect to the strategic lines analysed so far and to
derive constraints on which lines are more robust concerning our
present ignorance of the real situation.

\begin{figure}
\resizebox{\hsize}{!}
   {\psfig{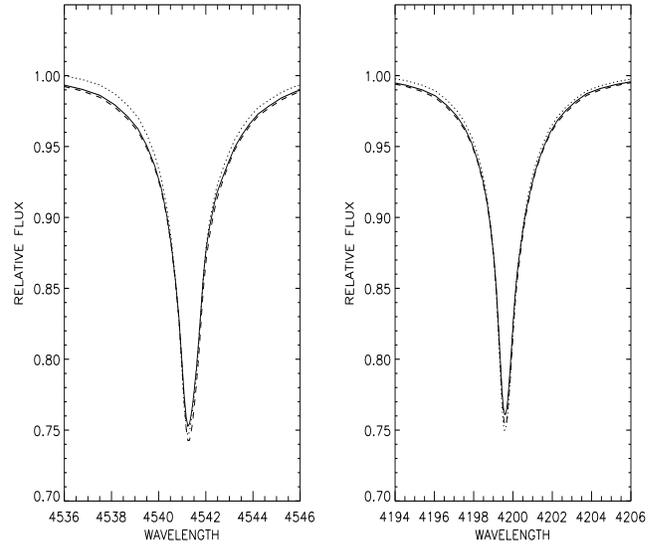}}
\caption[]{Line profiles of \HeII\,4541 (left) and \HeII\,4200 (right)
for the simulation of line-blocking (fully drawn), 
of detailed-balance (dashed), and for the ``standard'' model
(dotted). Calculations were made for the final model of
HD\,14\,947.}
\label{heiib}
\end{figure}

Fig.~\ref{heiib} shows the profiles of
\HeII\,$\lambda\lambda$\,4541, 4200 produced by the different
simulations and the ``standard'' model.  We see that the wings of the
\HeII\, lines become stronger in the simulations as a consequence of
the now reduced departure coefficients of the upper levels. However,
the effect is smaller in the \HeII\,$\lambda$4200 line (transition 4
$\rightarrow$ 11) than in the \HeII\,$\lambda$4541 (transition 4
$\rightarrow$ 9) because transitions involving higher levels are
weaker and form closer to the photosphere, so that the increased
source function is not so visible. Since \HeII\,4200 shows up to be
more stable, this is the preferred line in case of any discrepancy.
This is a change of criterion with respect to former analyses in our
group, but we prefer always to follow a single criterion that allows
us to understand physically changes in the derived parameters.  In
addition, we should mention that also \Ha\, becomes weaker, especially
in the blue wing, as a consequence of these effects in the overlapping
\HeII\, line. That is the reason why we prefer to fit the red wing in
case of difficulties like those in $\lambda$ Cep or HD\,15\,570 (and
in concert with the findings by Puls et al. (\cite{puls96}), who had
also to manipulate the \HeII\, departures predicted by unified models
if the wind was strong).  Finally, we point out that \HeI\, is not so
strongly affected if these lines are formed purely in the photosphere,
but, as has been shown by Santolaya--Rey et al.  (\cite{sr97}) they
are also influenced by any effects that modify the population of the
\HeII\, ground level (see Sect. 3.4.1 in Santolaya--Rey et al., where
this problem has already been discussed).

\section{Discussion}
\label{discu}

The first thing which is evident from the spectral analysis is the
increasing difficulty in fitting the spectra of the earliest types
with plane--parallel model atmospheres. In particular, for 50\,000~K
and above we were unable to determine the stellar parameters in the
same way as we did for stars in Paper I and for the rest of the
objects in the present study. This is true even for the relatively
large gravity star Cyg OB2 $\#$7, and was nearly the case for the
relatively cool star HD 15\,558. Plane--parallel analyses are still
useful, however, because they apply up to temperatures of 50\,000 K
and because they can be used as constraints for the analysis in the
larger parameter space demanded by more sophisticated models.

The temperature scale defined by plane--parallel, hydrostatic,
non-blanketed models is probably too hot (see Vacca et
al. \cite{vacc96}; Harries \& Hilditch \cite{hh98}). For example,
Hubeny et al. (\cite{hub98}) have shown that the same quality fit can
be achieved for 10 Lac with a line-blanketed model at 35\,000 K and
with the non-blanketed model used in Paper I at 37\,500 K (all other
parameters remaining the same). 
The effect can be even larger for the
stars analysed here, since the line-blocking effects in
plane--parallel, hydrostatic models move the stars towards higher
temperatures.

In the spherical, non-hydrostatic models we have used here,
line-blocking has been simulated, and we have found that it has the
same effect as to keep the \HeII\, resonance lines in detailed balance
(which were already in detailed balance in the plane--parallel models,
so that they did not show the influence on the \HeII~ lines of the
Pickering series we observe in these spherical models). This effect
had important consequences in our analyses. It led us to change the
temperature criterion and adopt \HeII~$\lambda$4200 as the main line
for the fit, which resulted in lower temperatures than if we had
adopted the \HeII~$\lambda$4541 in those cases in which we could not
fit both lines simultaneously.  The simulations indicate that the
reduction of the strong upward rates of the \HeII~ resonance lines
(whatever the real physical cause) will help to bring both lines into
agreement (although at the moment we cannot say whether it will bring
them completely into agreement). It also suggested to us that we
should fit the red wing of \Ha\, whenever the fit to the whole line
was impossible. What could be important for the future is that we have
shown that \HeII~$\lambda$4686 is also affected and its fit is highly
improved (although still qualitatively) when the resonance lines of
\HeII\, are kept in detailed balance, or the departure coefficient of
the second level is kept below its detailed balance value through
additional background opacity.

The plane--parallel spectroscopic masses are as usual lower than
evolutionary masses (see Paper I or Vacca et al., \cite{vacc96}).  The
new temperature criterion does not strongly influence the mass
discrepancy. In Paper I we showed that lowering the temperature (for
whatever reason) will not bring the masses into agreement. In the
present case, however, we find the mass discrepancy even for 
luminosity class V stars. This finding is contrary to the conclusion we
obtained in Paper I (and is in agreement with other authors; see, for
example, Vacca et al. \cite{vacc96}), namely that luminosity class V
stars do not show a mass discrepancy.  The reason for this apparent
contradiction is that the gravities we derive here are also low for
these luminosity class V stars. Thus the conclusion should be rather
that `high-gravity stars do not show a mass discrepancy,' where the
term `high' actually depends on the strength of the radiation field.

This indicates a problem of the hydrostatic models due to the intense
radiation pressure, and in fact larger gravities are derived when
using the spherical models with mass-loss, largely reducing the mass
discrepancy, which is now about 50$\%$. This result is even reinforced
when we realize that most temperatures are now lower due to sphericity
and a new temperature indicator.  An additional contribution came in
some cases from the fact that the wings of \Hg\, can be strongly
affected by wind contamination. In one case (HD\,15\,570) this
contamination is so strong that actual information about the gravity
from the wings of \Hg\, is lost. In all other cases, however, the
systematic effect that spectroscopic masses (without line-blanketing
or blocking) are lower than evolutionary ones (without mixing
mechanisms) is still present (note that formally the error bars
overlap, and it is only because the effect is systematic that we can
speak of a mass discrepancy).

The problem can be alternatively formulated using the escape
velocities. (Table 6 gives the escape velocities and related values).
Escape velocities have been derived using \g\, values uncorrected
for centrifugal forces, i.e., the {\it measured} values implicitely including
the centrifugal force acceleration (cf. Sect.~\ref{anap})
which determine the {\it effective} escape velocities.

\begin{table*}
\label{vesc}
\label{tabular}
\caption[ ]{$\Gamma$ values, terminal velocities and 
escape velocities (in km s$^{-1}$)
obtained using the spectroscopic (sp) and evolutionary (ev) masses.
In both cases we have taken into account the effect of the
centrifugal force in reducing the escape velocity.}
\begin{flushleft}
\begin{tabular}{llllrcrc} 
\hline
Star & \Vi & $\Gamma(sp)$ & $\Gamma(ev)$ & $v_{\rm esc}(sp)$ & 
\Vi / $v_{\rm esc}(sp)$ & $v_{\rm esc}(ev)$ & 
\Vi / $v_{\rm esc}(ev)$ \\
\hline
Cyg OB2 \#7& 2900 & 0.71 & 0.36 &  586 & 4.95 & 1231 & 2.36 \\
HD 15\,570 & 2600 & 0.31 & 0.43 & 1149 & 2.26 &  882 & 2.95 \\
HD 15\,629 & 3000 & 0.45 & 0.28 &  781 & 3.84 & 1148 & 2.61 \\
HD 15\,558 & 2800 & 0.46 & 0.43 &  981 & 2.85 & 1039 & 2.69 \\
HD 14\,947 & 2400 & 0.32 & 0.25 &  789 & 3.04 &  943 & 2.55 \\
HD 210\,839& 2250 & 0.33 & 0.31 &  734 & 3.07 &  778 & 2.89 \\
HD 5\,689  & 2500 & 0.40 & 0.10 &  405 & 6.17 & 1013 & 2.47 \\
\end{tabular}
\end{flushleft}
\end{table*}

\begin{figure}
{\psfig{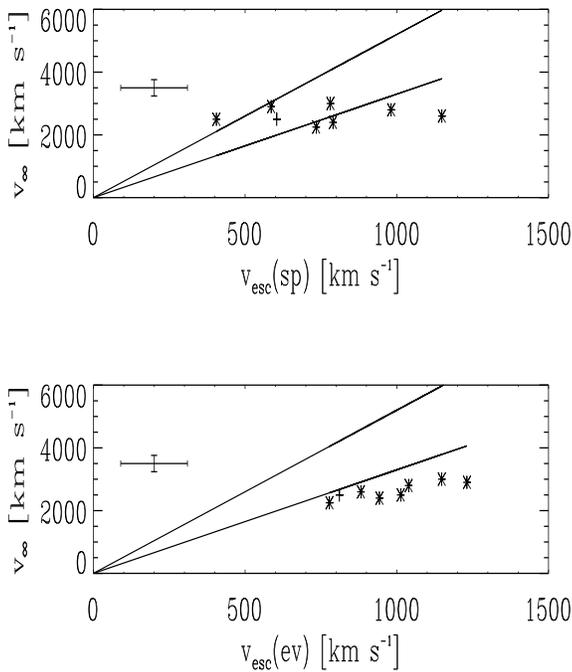}}
\caption[]{The escape velocities obtained from spectroscopic (above) 
and evolutionary (below) masses, against the wind terminal velocities,
all in km s$^{-1}$. The lines have the slopes predicted by theory
for $\alpha$ values of 0.6 and 0.7, e.g., 3.3 and 5.2. Plus signs
mark the position of HD\,5\,689 if this star is assigned 
a magnitude of --5.78, typical of a O6 III star. Typical error bars
have been plotted in the upper right corner. The abscissa error of
Cyg OB2$\#$7 in the upper plot is twice the corresponding error bar, due to
its high $\Gamma(sp)$ value}
\label{vescfig}
\end{figure}

Fig.~\ref{vescfig} shows the correlation between escape velocities
and terminal wind velocities. We see that the diagram using
evolutionary masses shows a good linear correlation between both
quantities, whereas that with the spectroscopic masses shows a weaker
correlation. However, the last diagram is in better agreement with the
theory of radiatively driven winds.  This theory predicts for the
O-star domain (i.e., if the force-multiplier parameter $\delta$ is
small, cf. Friend \& Abbott \cite{FA86}; Kudritzki et
al. \cite{kud89})
$$v_{\rm \infty} \simeq 2.24 {\alpha\over{1-\alpha}} v_{\rm esc},$$
with $v_{\rm esc}$ the escape velocity and $\alpha$
one of the line force multiplier parameters (the coefficient
in the exponent of the line strength distribution function).
With typical values for OB stars of $\alpha$= 0.6--0.7 we
obtain values of 3.3--5.2 for the ratio of
terminal-to-escape velocity. This range of values
is in agreement with those found here for the spectroscopic
masses (see Table~\ref{vesc}; 
note that the only point deviating from this range 
corresponds to HD\,15\,570, whose gravity and spectroscopic mass
are very uncertain, and that HD\,5\,689, which is the
leftmost point in the upper part of Fig.~\ref{vescfig}, would
lie in the middle of the range if the absolute magnitude of 
a luminosity class III object were assumed). 

On the other hand, we derive very low ratios between terminal and
escape velocities when using evolutionary masses (see
Table~\ref{vesc}), with an average ratio of 2.57, corresponding to
$\alpha$= 0.53 (which would be very low for the considered spectral
range).  Thus, Fig.~\ref{vescfig} seems to indicate that the
evolutionary masses are {\em systematically} too large (via the
corresponding escape velocities), whereas the spectroscopic ratios are
closer to the theoretically expected range.  In Fig. 29 we can see
this result, already contained in former papers (Groenewegen et
al. \cite{groe89}; Lamers \& Leitherer \cite{ll93}), from a slightly
different point of view. Here we have plotted the ratio of
evolutionary-to-spectroscopic mass versus the ratio of
spectroscopic-to-evolutionary $\alpha$. We see that, except for the
odd case HD\,5\,689, there is a strong correlation, indicating the
relation between mass discrepancy and our present knowledge of
radiatively driven winds.  (Note moreover that HD\,5\,689 would
perfectly fit into the correlation when an absolute magnitude of -5.78
is assumed, as appropriate for an O6 III star, but note also that the
terminal velocity for this star was derived from the spectral
type--terminal velocity relation from Haser \cite{has95}, as was that
of Cyg OB2$\#$7).  We should stress that we adopt a $\beta$--law for
the wind velocity and determine then the $\alpha$'s directly from the
derived relation between \Vi and escape velocity. Thus the agreement
of the spectroscopic $\alpha$'s with the predictions of the
radiatively driven wind velocity results in mutual support.

At present it is still unclear whether the inconsistency found is due
to some physical effects that should be incorporated into the
evolutionary or into the atmospheric models, although the former are
perfectly able to explain the discrepancy, at least qualitatively,
when introducing mixing effects (Heger \cite{heger98}) that could also
affect surface abundances.

\begin{figure}
\label{alfarat}
{\psfig{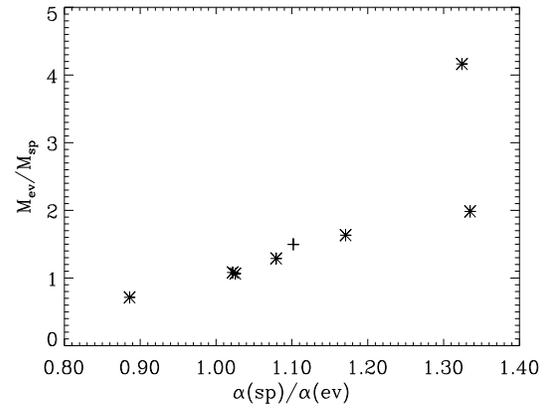}}
\caption[]{The ratio $\alpha$(sp)/$\alpha$(ev) against the mass ratio
\Mev/\Ms. Asterisks refer to the values quoted in Table 3 (in
particular, the asterisk in the upper right corner corresponds
to HD\,5\,689), whereas
the cross marks the position of HD\,5\,689 if this star is assigned 
a magnitude of --5.78, typical of an O6 III star.}
\end{figure}

Spherical models with mass-loss do not contribute to reduce the helium
abundances found with plane--parallel models. In previous analyses
(Herrero et al. \cite{h95}; Israelian et al. \cite{isr99}) in which we
used \HeI~ $\lambda\lambda$4387, 4922 and \HeII~ $\lambda$4541 as
temperature indicators, the derived helium abundances were similar.
As a result of the new indicator, the helium abundances had now to
increase. However, we see that the stars of larger gravity (HD 15\,558
and HD 15\,629) do not seem to show He overabundances.  Normal
supergiants lie between the former and the fast rotators, that show
the largest overabundances, in agreement with results from Paper I and
the possibility that rotation plays a fundamental role in the chemical
evolution of single stars. However, it is not clear whether a
difference exists in the mass discrepancy between the rapid rotators
and the other stars, or a correlation between the mass and helium
discrepancies, that could indicate an overluminous evolution, as
predicted by evolutionary rotating models (see Langer \& Heger
\cite{lanheg98}, Maeder \cite{mae98} or Meynet \cite{mey98} and
references therein). We should stress here that we do not consider HD
5\,689 as really showing such a large mass discrepancy as it appears
to do in Tables 2 and 3, but that it is a problem of the stellar
classification (or a problem of assigning the star to the Cas OB7
association).

Derivation of the mass--loss rates has been hindered by our finding of
the inconsistency between \Ha\, and \Hg\, in those cases in which the
wind is particularly strong.  Thus we have found that both lines can
demand mass--loss rates that differ up to a factor of two. We prefer
the values given by \Ha because it is more sensitive to mass-loss, and
attribute the problem to difficulties in describing the wind in the
transition zone.  IR observations and analyses giving information
about this zone should help in the future to solve the problem.

From the three stars for which we had some information about
mass--loss rates derived from radio emission, only one (for which in
addition only an upper limit from radio fluxes is available) shows
agreement between the \Ha\, and radio values. For the other two, radio
mass--loss rates are a factor of three to four lower than \Ha\,
mass--loss rates (and thus \Hg\, values would be in between).
However, several facts should be taken into account before claiming
that there is a contradiction between both sets of values.  First, the
concerned stars are only two particular, rather special cases (one,
HD\,15\,570, is a very extreme O star, and the other, HD\,210\,839, is
a strong non-radial pulsator, rapid rotator with a probably
non--spherically symmetric wind,
cf. Sect.~5.2); 
secondly, what we actually obtain from the observations are {\rm Q}
values, proportional to ($\Mdot/({\rm R} {\rm v}_\infty)^{3/2}$) for
\Ha , and to ($\Mdot/({\rm R}^{3/2} {\rm v}_\infty)$) for radio and
thus the derived values depend on the particular set of chosen stellar
parameters; and thirdly, also the proportionality constants in the
case above depend on model details (for example, Lamers \& Leitherer
\cite{ll93}, assume that He is doubly ionized in the wind of
HD\,210\,839 between 10 and 100 stellar radii, whereas our
calculations indicate that from 28 stellar radii upwards, He is single
ionized, which would have an immediate effect on the derived radio
mass--loss rates, increasing them).  Thus, we cannot state that there
is a general problem (or even a particular one), or decide which value
we should prefer for each object.

Having derived all parameters we can obtain the modified wind momenta
of all stars for which the radiation-driven wind predicts a tight
correlation with luminosity. Our results are given in Table 3 and
plotted in Fig.~\ref{wlrfig}, together with the data given by Puls et
al. (\cite{puls96}) for Galactic supergiants.  We see that our points
for supergiants agree well with theirs (we have discussed the
differences in the individual studies). We have also plotted the
regression lines derived from the supergiants from Puls et al.  and
from all the supergiants (the former and ours) to stress this
point. We see that both regression lines are quite parallel,
indicating that the new values do not significantly change the known
WLR.

\begin{figure}
{\psfig{figure=h1606.f30,width=8.5cm,height=8.0cm}}
\caption[]{The wind momentum--luminosity relation for the stars in our
sample. The abcissa is \lL~ and the ordinate is the logarithm of the
modified wind momentum (MWM),
\lMWM . Asterisks represent OB supergiants from
Puls et al. (\protect\cite{puls96}). Open diamonds, squares and triangles  
are respectively OB supergiants, giants and dwarfs from the present 
work. The positions of HD\,14\,947 and HD\,210\,839 in both
studies are joined by a line, where the major difference results from the
different effective temperatures and helium-abundances attributed to the
objects.}
\label{wlrfig}
\end{figure}

\section{Conclusions and future work}
\label{conc}

We have presented the spectra of seven Galactic luminous O stars, that
have been analysed by means of non-LTE, plane--parallel, hydrostatic
model atmospheres, including line-blocking, and by means of spherical,
mass-losing models. We have used these analyses to study some
additional effects (physical as well as due to applied approximations)
that have an influence on the results.

We have shown that line-blocking has to be taken into account when
analysing very hot stars, in plane--parallel as well as in spherical
models with mass-loss. Line-blocking indirectly affects the
spectral-line diagnostics through changes in the level populations of
He. Although for the spherical models with mass-loss we only made
simulations, this is clearly one direction for future model
improvements. In particular, we obtained very interesting results
concerning the \HeII\,$\lambda$4686 line and the blue wing of
\Ha\,(actually, concerning the corresponding \HeII\,blend).

One of the conclusions of the present paper is that around 50\,000 K
(the exact value depending on gravity) we reach the limit of
applicability of plane--parallel, hydrostatic models to real massive
stars. This conclusion can be further refined, if we take into account
the fact that line-blanketed models could give lower temperatures, or
if we consider that spherical models with mass-loss will demand larger
gravities. The final result would be models less affected by radiation
pressure, larger masses and a cooler temperature scale.

Spherical models with mass-loss are thus needed to analyse these
stars.  We had to correct the gravities of all stars (except HD
5\,689) when using these models, sometimes with increments as large as
0.25 dex.  Part of the differences in the stellar parameters derived
from both sets of models are due to a change in the preferred
temperature indicator.

Our spherical models with mass-loss are also still not free from
internal inconsistencies. For some models with strong winds, we see
that \HeII\,$\lambda\lambda$4200, 4541 give different temperatures,
although we expect that inclusion of line-blocking will reduce this
discrepancy, and will strongly improve the fits of
\HeII\,$\lambda$4686 and the blue wing of \Ha .  We have also found
that for stars with strong winds there is a difference in the
mass--loss rates derived from \Ha\, and \Hg , that can reach a factor
of two. We do not find good agreement with the two cases for which we
have mass--loss rates from radio fluxes, although it is not possible
to derive any firm conclusion from this fact.

The helium and mass discrepancies found here are in agreement with the
findings of Paper I. The only difference is that we find a mass
discrepancy also for luminosity class V stars, but this is actually
misleading. What we find is that the luminosity class V stars analysed
here have gravities lower than usual for their spectral
classification.  The conclusion of Paper I should then be changed to
state that we do not find the mass discrepancy for stars with high
gravities.  At least a part of this effect is attributed to the fact
that the intense radiation field affects the wings of the Balmer lines
also for these stars.  An additional possibility is that they actually
never reach the zero-age main sequence (see, for example, Hanson
\cite{han98}).

Spherical models with mass-loss largely reduce the mass
discrepancy.  Without solving it completely, the problem lies now in
the systematic trend of spectroscopic values to be lower than
evolutionary ones, but most of the individual values now agree within
the formal error bars.  We have also shown that the spectroscopic
masses agree better with the predictions from the radiatively driven
wind theory, because they give ratios of the terminal wind velocity to
the escape velocity (or equivalently, values of the $\alpha$ line
force parameter) that are in the range predicted by that theory.

From the point of view of individual stars, we have analysed some
of the most massive and luminous stars in the Milky Way. We have found
that three of them (Cyg OB2 $\#$7, HD\,15\,570 and HD\,15\,558) have
particularly large initial masses, around or in excess of 100 \Msun\
(depending on the technique used for the mass derivation).  Cyg
OB2$\#$7, HD\,14\,947 and HD\,210\,839 have so extreme mass--loss
rates that the wings of \Hg\, are strongly affected, and in
HD\,15\,570 the wind is so strong that the exercise of deriving the
gravity from the wings of \Hg\, results in highly uncertain values.
On the other hand, we find that HD 5\,689, the less luminous object in
our sample, could have been wrongly assigned to Cas OB7, and might be
a runaway star.

\acknowledgements{We would like to thank our anonymous referee
for many valuable comments and suggestions. AH wants to acknowledge
support for this work by the spanish DGES under project PB97-1438-C02-01.}

\end{document}